\documentclass[twocolumn,superscriptaddress,preprintnumbers]{revtex4}
\usepackage{graphicx}
\usepackage[figuresright]{rotating}
\usepackage{psfrag}
\newcommand{\beq}{\begin{equation}}
\newcommand{\eeq}{\end{equation}}
\newcommand{\be}{\begin{eqnarray}}
\newcommand{\ee}{\end{eqnarray}}
\begin{document}
\title{Chiral dynamics and the growth of the nucleon's gluonic 
transverse size at small $x$}
\author{M.~Strikman} 
\affiliation{Pennsylvania State University,
University Park, PA 16802, U.S.A.}
\author{C.~Weiss}
\affiliation{Institut f{\"u}r Theoretische Physik,
Universit{\"a}t Regensburg, D--93053 Regensburg, Germany}
\begin{abstract}
We study the distribution of gluons in transverse space in the nucleon 
at moderately small $x$ ($\sim 10^{-2}$). At large transverse distances 
(impact parameters) the gluon density is generated by the ``pion cloud'' 
of the nucleon, and can be calculated in terms of the gluon density in 
the pion. We investigate the large--distance behavior in two different 
approaches to chiral dynamics: {\it i)}~phenomenological soft--pion 
exchange, {\it ii)}~the large--$N_c$ picture of the nucleon as a classical 
soliton of the pion field, which corresponds to degenerate $N$ and $\Delta$ 
states. The large--distance contributions from the ``pion cloud'' cause a 
$\sim 20\%$ increase in the overall transverse size of the nucleon if $x$ 
drops significantly below $M_\pi / M_N$. This is in qualitative agreement 
with the observed increase of the slope of the $t$--dependence of the 
$J/\psi$ photoproduction cross section at HERA compared to fixed--target 
energies. We argue that the glue in the pion cloud could be probed directly 
in hard electroproduction processes accompanied by ``pion knockout'', 
$\gamma^\ast  + N \; \rightarrow \; \gamma \; 
(\mbox{or}\;\; \rho, \, J/\psi ) \; + \pi + N'$, where the transverse 
momentum of the emitted pion is large while that of the outgoing nucleon 
is restricted to values of order $M_\pi$.
\end{abstract}
%
%
\maketitle
\section{Introduction}
Hard scattering processes induced by (real or virtual) photons are an
important source of information about the structure of the nucleon.
The parton model description of such processes distinguishes a
``longitudinal'' direction, defined by the 3--momentum of the incoming
photon, and the ``transverse'' plane perpendicular to it.  The
structure functions of inclusive deep--inelastic scattering are
proportional to the parton densities, describing the distribution of
partons with respect to longitudinal momentum of the parent nucleon;
they do not carry any information about the distribution of partons in
the transverse plane. In this sense, inclusive deep--inelastic
scattering provides us with a 1--dimensional image of the
nucleon. Much more detailed information can be obtained from exclusive
processes, in which one measures amplitudes with a non-zero momentum
transfer between the initial and final nucleon, 
$\Delta = p' - p \neq 0$, with $t \equiv \Delta^2 < 0$. These include
hard electroproduction processes such as deeply--virtual Compton
scattering and meson production, or diffractive photoproduction of
heavy quarkonia ($J/\psi, \Upsilon$), which probes the gluon
distribution in the nucleon. Such measurements can give information
also about the spatial distribution of partons in the transverse
plane, thus providing us with a 3--dimensional image of the nucleon.

A precise formulation of the notion of a spatial distribution of
partons in the transverse plane is possible within the formalism of
generalized parton distributions (GPD's), which parametrize the
non-forward matrix elements ($p' \neq p$) of twist--2 QCD light--ray
operators between nucleon states
\cite{Muller:1998fv,Ji:1997nm,Radyushkin:1997ki,Collins:1996fb}.  
Of particular interest is the ``diagonal'' limit of zero longitudinal
component of the momentum transfer, $\Delta^+ = 0$. The GPD's in this
case depend (in addition to the partonic variable, $x$) only on the
transverse component of the momentum transfer, ${\bf\Delta}_\perp$,
with $-{\bf\Delta}_\perp^2 = t$, and the nucleon helicity--conserving
ones reduce to the usual polarized and unpolarized parton densities in
the limit $t \rightarrow 0$ (for this reason the $\Delta^+ = 0$ GPD's
are also referred to as ``non-forward parton densities''
\cite{Radyushkin:1998rt}).  The Fourier transform of these functions
with respect to ${\bf\Delta}_\perp$ then defines functions of a
2--dimensional coordinate variable, ${\bf b}$, which can be
interpreted as the spatial distributions of partons with longitudinal
momentum fraction $x$ in the transverse plane (``impact
parameter--dependent parton distributions'') \cite{Burkardt:2002hr}.
The integral of these ${\bf b}$--dependent distributions reproduce the
total densities of partons for given $x$, and they can be shown to
satisfy positivity conditions locally in ${\bf b}$
\cite{Pobylitsa:2002iu}.  This ``mixed'' momentum and coordinate
representation provides a vivid 3--dimensional (more precisely, 
$1 + 2$--dimensional) picture of the partonic structure of the
nucleon, which naturally lends itself to the visualization of
polarization effects, including orbital angular momentum
\cite{Burkardt:2002hr}.  On the phenomenological side, it has been
speculated that this representation could provide a simple explanation
of the transverse single spin asymmetries observed in semi-inclusive
deep--inelastic scattering \cite{Burkardt:2002ks}.

It should be noted that the ``diagonal'' GPD's ($\Delta^+ = 0$) are
not directly observable, since hard exclusive processes generally have
skewed kinematics. However, in a number of important cases the
kinematics is effectively not far from the diagonal case. In
particular, this is the case for $J/\psi$ photoproduction, which
probes the generalized gluon distribution in the nucleon; the ratio of
the longitudinal momentum fractions of the two gluons is of the order
$x_1/x_2 \sim 2$ \cite{Frankfurt:1997fj}. Furthermore, for $\Upsilon$
production \cite{Martin:1999rn}, and generally for electroproduction
at large $Q^2$ and sufficiently small $x$ \cite{Frankfurt:1998yf}, the
gluon GPD's are dominated by evolution down from significantly larger
values of $x_1$ and $x_2$. Since the difference $x_1 - x_2$ is not
changed by the evolution, this implies that one is actually probing
the $t$--dependence of the nearly diagonal distributions at larger
$x_1$ and $x_2$ (the $t$--dependence is not affected by the
evolution).

An interesting aspect of the impact parameter representation of parton
distributions, which has been less emphasized so far, is that it
provides a useful new framework for developing dynamical models for
(generalized) parton distributions. The possibility to separate
contributions from different transverse distance scales allows one to
introduce the notion of a ``size'' of the nucleon, well--known from
elastic form factors (charge radii), in the discussion of generalized
parton distributions. Furthermore, one can incorporate the fundamental
fact that the long--distance behavior of strong interactions is
governed by chiral dynamics. Thus, one would expect that, in a certain
range of $x$, the parton distributions at transverse distances of
order $1/M_\pi$ can be attributed to the ``pion cloud'' of the
nucleon, and can be computed from first principles in terms of the
known parton distributions in the pion. This would provide
model--independent constraints for the distributions of partons in the
transverse plane. Moreover, the partons in the pion cloud may lead
directly to observable effects in hard processes sensitive to large
transverse distances.

In this paper we study the impact parameter dependence of the
nucleon's gluon density from a phenomenological perspective.  We
concentrate on the region of moderately small $x$ ($\sim 10^{-2}$),
for which the $t$--dependence of the gluon GPD (the so-called
``two--gluon form factor'' of the nucleon) has been extracted from a
number of electro--/photoproduction experiments, see
Ref.\cite{Frankfurt:2002ka} for a compilation.  The investigation
consists of three parts.  First, we derive the asymptotic behavior of
the impact parameter--dependent distribution at large 
$b \equiv |{\bf b}|$. In this limit the gluon distribution is
dominated by soft--pion exchange contributions and calculable in terms
of the gluon density in the pion.  The asymptotic behavior at large
$b$ can be stated in the form of a DGLAP--type convolution of the
gluon density in the pion with a $b$--dependent distribution of pions
in the nucleon, which is completely determined by chiral dynamics and
exhibits a Yukawa-like falloff at large $b$. Second, we investigate
the contribution of large transverse distances, $b \gtrsim 1/M_\pi$,
to the average transverse size of the nucleon, 
$\langle b^2 \rangle$. The latter is proportional to the slope of the
$t$--dependent gluon GPD at $t = 0$, and thus directly observable. We
find that the large--distance contributions lead to a finite increase
in the transverse size of the nucleon if $x$ drops significantly below
$\sim M_\pi / M_N$. Using a simple two--component picture, we estimate
the magnitude of the increase at roughly $20\%$ of the total
transverse size of the nucleon. Such an increase is indeed observed
when comparing the slope of the $t$--dependence of the $J/\psi$
photoproduction cross section at HERA with that at fixed--target
energies \cite{Frankfurt:2002ka}.  Third, we ask how the glue in the
pion cloud could be probed directly in hard exclusive processes. A
promising candidate is hard electroproduction of photons 
(or $\rho, J/\psi$ mesons) accompanied by the ``knockout'' of an
additional pion, $\gamma^\ast + N \; \rightarrow \; \gamma \;
(\mbox{or}\;\; \rho, \, J/\psi ) \; + \pi + N'$, such that the
transverse momentum of the pion is large while that of the outgoing
nucleon is restricted to values of order $M_\pi$.

It is worth noting that there is an interesting connection between the
increase of the transverse size of the nucleon due to the pion cloud
described here, and the so-called shrinkage of the diffractive slope
in soft physics. In soft physics the shrinkage is usually interpreted
as due to Gribov diffusion of the impact parameters in the partonic
ladder, see Ref.\cite{Gribov:jg} for a pedagogical discussion. The
phenomenon we discuss here is essentially related to the diffusion in
the first rung of the ladder, which is delayed as compared to soft
physics by selection of relatively large--$x$ gluons which are missing
in the pion cloud.

A crucial element in our approach is the impact--parameter dependent
``parton distribution'' of pions in the nucleon at large transverse
distances, $b \gtrsim 1/M_\pi$. We show that this concept is
meaningful only for pion momentum fractions $y < M_\pi / M_N$, where
the pion virtuality is of order $M_\pi^2$. We calculate this
distribution in two different approaches to chiral dynamics ---
soft--pion exchange with a phenomenological pion--nucleon coupling,
and the large--$N_c$ picture of the nucleon as a classical soliton of
the pion field. We explicitly demonstrate the equivalence of the two
approaches. The key to this is the realization that in the
large--$N_c$ limit the nucleon and the Delta resonance are degenerate,
so the large--$N_c$ calculation based on the classical soliton
corresponds to soft--pion exchange with $N$ and $\Delta$ intermediate
states included on the same footing. We also comment in general on the
relation of pion cloud contributions at large impact parameters to the
$1/N_c$--expansion of parton distributions \cite{Diakonov:1996sr}. Our
use here of the large--$N_c$ limit is completely general, and not
related to any particular dynamical model used to generate stable
soliton solutions.  Nevertheless, the results obtained here provide
useful constraints for calculation of parton distributions within
specific large--$N_c$ models, such as the chiral quark--soliton model
\cite{Petrov:1998kf}.

Superficially, our approach bears some resemblance to the well--known
pion cloud model used to explain the observed flavor asymmetry of the
sea quark distributions in the proton, $\bar d(x) > \bar u(x)$
\cite{Sullivan:1971kd}, see {\it e.g.}\ Refs.\cite{Kumano:1997cy} for
recent reviews. There is, however, an important conceptual difference
in that we invoke the notion of pion exchange only at large transverse
distances, $b \gtrsim 1/M_\pi$, where the contributions from $\pi N$
configurations in the nucleon wave function are distinct from those of
average configurations and thus physically meaningful. This way of
defining the ``pion cloud'' contributions naturally avoids a number of
problems which have plagued the traditional pion cloud model, see 
{\it e.g.}\ the discussion in Ref.\cite{Koepf:1995yh}. In our approach
the pions are almost on-shell (the virtuality is of order $M_\pi^2$),
we do not need cutoffs at the pion--nucleon vertices, and the results
are independent of the form of pion--nucleon coupling (pseudoscalar or
axial).

In the present investigation we concentrate on the unpolarized gluon
distribution, the $t$--dependence of which is well--known in the
regions of $x$ covered by $J/\psi$ photoproduction at fixed--target
and collider energies \cite{Frankfurt:2002ka}.  The approach developed
here can easily be extended to the quark distributions (singlet and
non-singlet), as well as to polarized quark and gluon
distributions. It can also be applied to the diagonal (zero
skewedness) limit of GPD's with a helicity flip between the nucleon
states, which have no correspondence in the parton distributions of
inclusive deep--inelastic scattering.

The study of the transverse size of the gluon distribution in the
nucleon is important also for a number of other problems, not directly
related to exclusive processes. Knowledge of the transverse size of
the gluon distribution at moderate $x$ is required for modeling the
initial conditions for non-linear QCD evolution equations describing
the saturation of parton densities at very small $x$
\cite{Jalilian-Marian:1996xn,Balitsky:1995ub,Kovchegov:1999ua}.  The
transverse size of the hard component of the nucleon wave function is
also an important parameter in modeling hadron--hadron collisions at
LHC energies \cite{FELIX}.

This paper is organized as follows. In Section~\ref{sec_npd} we review
the basic properties of the impact parameter representation of the
gluon density, starting from its covariant definition as the Fourier
transform of the $t$--dependent non-forward gluon density.  In
Section~\ref{sec_large} we derive the asymptotic behavior of the gluon
density in the nucleon at large impact parameters as due to soft--pion
exchange, and discuss its dependence on $x$. 
In Section~\ref{sec_largenc} we address the same problem from the
point of view of the large--$N_c$ limit, where the nucleon is
described as a classical soliton of the pion field, and demonstrate
the equivalence of this picture to soft--pion exchange including Delta
resonance contributions.  In Section~\ref{sec_size} we show that the
pion cloud contribution at $b \gtrsim 1/M_\pi$ produces a finite
increase of the overall transverse size of the nucleon, 
$\langle b^2 \rangle$, if $x$ drops significantly below $M_\pi / M_N$,
and estimate the magnitude of the increase in a simple two--component
picture. In Section~\ref{sec_knockout} we explore the possibility of
measuring the glue in the pion cloud directly by way of
electroproduction of photons or vector mesons at small $t$ accompanied
by large--angle pion production (``pion knockout'').  Our conclusions
and possible generalizations of the approach proposed here are
summarized in Section~\ref{sec_conclusions}.
\section{Impact parameter representation of the gluon density}
\label{sec_npd}
The matrix element of the twist--2 QCD gluon operator between nucleon
states of different momenta exhibits two Dirac structures (nucleon
helicity--conserving and --flipping), and is parametrized as
\be
&& 2 \int_{-\infty}^\infty
\frac{d\lambda}{2\pi} e^{i\lambda x (Pn)} n^\alpha n^\beta
\; \langle N(P + \Delta / 2) |
\nonumber \\
&\times& G_\alpha^{A\gamma} (-\lambda n/2) \,
G^A_{\gamma\beta} (\lambda n/2) \;
|N(P - \Delta / 2) \rangle
\nonumber \\
&=&
x H_g (x, t) \; \bar u' \hat n u 
\; + \; \frac{x E_g (x, t)}{2 M_N} \;
\bar u' \sigma_{\mu\nu} \Delta^\mu n^\nu u .
\label{H_g_E_g_def}
\ee
Similar definitions apply to the matrix elements of the twist--2 quark
operators, see Ref.\cite{Ji:1997nm}.  Here, $n$ is a light--like
four--vector, $n^2 = 0$, whose normalization is arbitrary, and
$G_{\alpha\beta}^A$ denotes the gluon field strength (here and in the
following, a gauge link between the fields is implied but will not be
written). The incoming and outgoing nucleon four--momenta are
expressed in terms of the average momentum, $P$, and the momentum
transfer, $\Delta$. The nucleon spinors, $u \equiv u (P - \Delta /2)$
and $\bar u' \equiv \bar u (P + \Delta /2)$ (we suppress the helicity
labels for brevity) are normalized according to 
$\bar u u = \bar u' u' = 2 M_N$, and our conventions are 
$\hat n \equiv n^\mu \gamma_\mu$ {\it etc.}, and 
$\sigma_{\mu\nu} \equiv (1/2) [\gamma_\mu , \gamma_\nu ]$.
The functions $H_g$ and $E_g$ in Eq.(\ref{H_g_E_g_def}) depend on the
partonic variable, $x$, and the invariant momentum transfer, 
$t \equiv \Delta^2 < 0$, and are referred to as the non-forward gluon
densities of the nucleon. It is implied here that
\beq
\xi \;\; \equiv \;\; -2 \, \frac{(\Delta n)}{(P n)} \;\; = \;\; 0 ;
\label{nonskewed}
\eeq
in the general case the functions $H_g$ and $E_g$ would depend also on
the ``skewedness'' parameter, $\xi$.  Finally, the functions $H_g$ and
$E_g$ depend implicitly also on the normalization point of the QCD
operator; this dependence is governed by evolution equations. 
In what follows we have in mind a typical scale for hard processes,
say $\sim 4 \, {\rm GeV}^2$, which will not be exhibited explicitly.

The non-forward gluon densities are defined on the interval $-1 < x < 1$.
Due to $C$--invariance they are odd functions of $x$,
\beq
H_g (-x, t) \;\; = \;\; -H_g (x, t),
\eeq
and similarly for $E_g (x, t)$. At $t = 0$ the function $H_g$ for $x > 0$ 
reduces to the usual gluon distribution in the nucleon (as defined 
{\it e.g.}\ in Ref.\cite{Collins:1981uw})
\beq
H_g (x, t = 0) \;\; = \;\; g(x)
\hspace{2em} \mbox{for $x > 0$.}
\label{NPD_forward}
\eeq
The $n$'th moments in $x$ of the functions $H_g (x, t)$ and $E_g (x, t)$ 
define the form factors of the local twist--2 spin--n gluon operators, 
which arise from the Taylor expansion of the non-local light-ray operator 
in Eq.(\ref{H_g_E_g_def}) in powers of the separation, $\lambda$.
In particular, the second moments are related to the form factors of the 
gluonic part of the QCD energy--momentum tensor \cite{Ji:1997nm}. 
Note that with our conventions
\beq
\frac{1}{2} \int_{-1}^1 dx \, x H_g (x, t = 0) \;\; = \;\; 
\int_0^1 dx \, x g(x) \;\; = \;\; A_g ,
\eeq
where $A_g$ parametrizes the traceless part of the expectation value 
of the gluonic part of the QCD energy--momentum tensor in the nucleon,
\be
\langle N(P) | \; G_\alpha^{A\gamma} (0) G^A_{\gamma\beta} (0) \; 
| N(P) \rangle
&=& 2 \, A_g \; P_\alpha P_\beta 
\nonumber \\
&+& \mbox{trace terms} , 
\ee
and can be interpreted as the total momentum fraction carried by gluons
in the parton model.

For the discussion of the $t$--dependence of the non-forward gluon density
it is convenient to define the so-called two--gluon form factor
of the nucleon \cite{Frankfurt:nc,Frankfurt:2002ka}
\beq
\Gamma (x, t) \;\; \equiv \;\; \frac{H_g (x, t)}{H_g (x, t = 0)}
\hspace{3em} (x > 0),
\label{twogluon_def}
\eeq
which is normalized to $\Gamma (x, t = 0) = 1$. This function can be
interpreted as the form factor describing the distribution of partons
with given longitudinal momentum fraction, $x$, and can be directly
compared with the well-known elastic form factors of the nucleon
(electromagnetic, axial).

The non-forward gluon densities are invariant functions, not
associated with any particular reference frame. However, as with the
invariant elastic form factors, it is possible to give an
interpretation of these functions as Fourier transforms of certain
``charge densities'' in a special frame in which the momentum transfer
has only spatial components \cite{Burkardt:2002hr}. Choosing $n$ to
define the light-cone ``plus'' direction, $n^\mu = (1, 0, 0, -1)$,
\be
P^+ &\equiv& (nP) \;\; = \;\; P^0 + P^3 , \\
P^- &\equiv& P^0 - P^3 ,
\label{P_plus_def}
\ee
one goes to a frame in which ${\bf P}_\perp = 0$. The condition 
$(n\Delta) = 0$ implies $\Delta^+ = 0$, and the kinematical constraint
$(P\Delta) = 0$ requires that also $\Delta^- = 0$, {\it i.e.}, the
momentum transfer has only a transverse component, 
${\bf \Delta}_\perp$, with
\beq
{\bf \Delta}_\perp^2 \;\; = \;\; -t .
\label{Delta2_from_t}
\eeq
In this frame the non-forward density can be thought of as a function
of $x$ and the transverse vector ${\bf\Delta_\perp}$. The dependence on 
${\bf\Delta}_\perp$ can be represented as a Fourier integral over a
transverse coordinate variable (``impact parameter''), ${\bf b}$: 
\beq
H_g (x, -{\bf \Delta}_\perp^2 ) \;\; = \;\; \int d^2 b \; 
e^{i ({\bf b} {\bf \Delta}_\perp)}
\; g(x, b) ,
\label{impact_def}
\eeq
where $b \equiv |{\bf b}|$. The limiting relation (\ref{NPD_forward}) 
for the non-forward gluon density at $t = 0$ then implies that
\beq
\int d^2 b \; g(x, b) \;\; = \;\; g(x) .
\eeq
Thus, one may think of the function $g(x, b)$ as describing the
distribution of partons of given longitudinal momentum fraction $x$ in
the transverse plane, in the sense that the integral over the transverse 
plane gives the usual parton densities, depending only on $x$. 
In Ref.\cite{Burkardt:2002hr} the name ``impact parameter--dependent 
parton distribution'' was proposed for these functions.

An important feature of the impact parameter--de\-pen\-dent parton 
distributions is that they satisfy positivity conditions locally
in ${\bf b}$ \cite{Pobylitsa:2002iu}. In particular, it makes sense to 
compute averages over the transverse plane, with the function $g(x, b)$ 
acting as a positive definite weight. For instance, one can define 
the ($x$--dependent) average transverse size squared of the nucleon as
\beq
\langle b^2 \rangle
\;\; \equiv \;\; \frac{\int d^2 b \; g(x, b) \; b^2}
{\int d^2 b \; g(x, b)} ;
\label{b2_def}
\eeq
the integral in the denominator is just the total gluon density, 
$g(x)$. This average can also be expressed as the $t$--derivative of the 
non-forward gluon density, {\it viz.}\ the two--gluon form factor,
Eq.(\ref{twogluon_def}), at $t = 0$. Differentiating Eq.(\ref{impact_def}) 
twice with respect to the vector 
${\bf \Delta}_\perp$ at ${\bf \Delta}_\perp = 0$, and
making use of the relation (\ref{Delta2_from_t}), one obtains
\be
\langle b^2 \rangle &=&
\frac{4}{H_g (x, 0)} \; \left. 
\frac{\partial H_g (x, t)}{\partial t} \; \right|_{t = 0}
\nonumber \\
&=& 4 \, \left. \frac{\partial \Gamma (x, t)}{\partial t} \; \right|_{t = 0} .
\label{b2_from_slope}
\ee
The factor 4 here represents two times the dimension of the transverse
plane; it replaces the factor of 6 in the well--known relation between
the slope of the elastic form factors and the ``three--dimensional''
charge radii.
\section{Large transverse distances and the pion cloud}
\label{sec_large}
We now study the behavior of the nucleon's gluon distribution at large
impact parameters from a phenomenological point of view. For
understanding the physical mechanism governing the large--$b$
asymptotics one may start with the moments in $x$ of the
$t$--dependent nonforward gluon densities, {\it i.e.}, the nucleon
form factors of the local twist--2 spin--$n$ gluon operators, which
are functions only of the invariant momentum transfer, $t$.  Their
Fourier transforms with respect to ${\bf\Delta}_\perp$ 
(with $t = -{\bf\Delta}_\perp^2$) define the moments of the impact
parameter--dependent gluon distributions. The asymptotic behavior of
the latter at large $b$ is governed by the singularities in $t$ of the
$t$--dependent moments. Since the local twist--2 gluon operators are
$G$--parity--even, the singularity closest to $t = 0$ is the cut at 
$t > 4 M_\pi^2$ coming from two--pion exchange in the $t$--channel.
The threshold behavior of the imaginary part at 
$t \rightarrow 4 M_\pi^2 + 0$ dominates the large--$b$ behavior of the
moments of the $b$--dependent gluon distribution.

\begin{figure}[t]
\begin{center}
\psfrag{pi1}{{\Large $\pi$}}
\psfrag{mln2}{{\Large $-\lambda n/2$}}
\psfrag{ln2}{{\Large $\lambda n/2$}}
\psfrag{N1}{{\Large $N$}}
\psfrag{NDel}{{\Large $N (\Delta)$}}
\includegraphics[width=8.4cm,height=4.9cm]{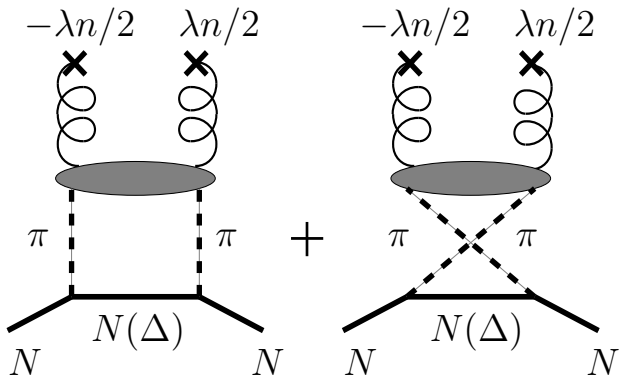}
\end{center}
\caption[]{The pion exchange contribution to the non-forward gluon
densities of the nucleon, Eq.(\ref{H_g_E_g_def}), determining the
leading asymptotic behavior of the impact parameter--dependent gluon
densities at large $b$.}
\label{fig_graphs}
\end{figure}

Having noted this, we may study the pion exchange contributions to the
non-forward gluon densities directly for the $x$--dependent functions,
{\it i.e.}, for the matrix element of the non-local gluon operator of
Eq.(\ref{H_g_E_g_def}), see Fig.~\ref{fig_graphs} \footnote{The soft
pion--nucleon Lagrangian contains an $N N \pi \pi$ vertex in addition
to the $N N \pi$ ``Yukawa'' coupling. Thus, in general, soft--pion
exchange contributions to the GPD's include also graphs in which the
two pions couple to the nucleon in the same point. However, these
graphs contribute to the GPD only in the region $-\xi < x < \xi$
[$\xi$ is the skewedness parameter, {\it cf.}\ Eq.(\ref{nonskewed})],
and can be dropped in the case $\xi = 0$ considered here.}. The dashed
lines indicate the pion, the solid line the nucleon (later we shall
consider also graphs with $\Delta$ intermediate states), and a
phenomenological pion--nucleon coupling is assumed at the vertices.
Before specifying how to interpret and evaluate the graphs of
Fig.~\ref{fig_graphs}, let us emphasize that we shall use them only as
a means to derive the large--$b$ asymptotics of the impact
parameter--dependent gluon distributions in the nucleon (or,
equivalently, the $t \rightarrow 4 M_\pi^2 + 0$ threshold behavior of
the imaginary part of the $t$--dependent non-forward gluon densities),
which is governed by exchange of {\it soft} pions, much like the
nucleon--nucleon interaction at large distances (Yukawa potential). We
shall not attempt to make sense of the resulting Feynman integrals in
regions where the pion virtuality is $\gg M_\pi^2$.

The blob in the graphs of Fig.~\ref{fig_graphs} represents the
non-forward gluon density in the pion. In the physical region $t < 0$
this function parametrizes the on-shell matrix element of the QCD
gluon operator between physical pion states,
\be
\lefteqn{
n^\alpha n^\beta \; \langle \pi^a (k + \Delta/2) |\;
G_{\alpha}^{A\gamma} (-\lambda n/2)
\, G^A_{\gamma\beta} (\lambda n/2) }
&& \nonumber \\
&\times& | \pi^b (k - \Delta / 2) \rangle
\nonumber \\
&=& \delta^{ab} \; (kn)^2
\int_{-1}^1 dz \; e^{-i z \lambda (kn)} \; z H_{g, \pi} (z, t) ,
\label{H_g_pi}
\ee
where again $(\Delta n) = 0$. The non-forward gluon density in the
pion $H_{g, \pi} (z, t)$ is an analytic function of $t$ which can be
continued to the unphysical region. In the soft--pion exchange
contribution to the distribution in the nucleon this function enters
at the two--pion threshold, $t = 4 M_\pi^2 > 0$, in the unphysical
region. By crossing symmetry, it could be related to the two--pion
gluon distribution amplitude ({\it i.e.}, the matrix element of the
same gluon operator between the vacuum and a two--pion state) in the
physical region for two--pion production; for a discussion of the
crossing relations see Refs.\cite{Polyakov:1999gs}.
In the following we shall neglect the $t$--dependence of the
non-forward gluon density in the pion, {\it i.e.}, we replace the
nonforward by the usual ($t = 0$) gluon density in the pion, 
$g_\pi (z)$:
\beq
H_{g, \pi} (z, t) \;\; \rightarrow \;\; H_{g, \pi} (z, t = 0) 
\;\; = \;\; 
\left\{\begin{array}{rl} g_\pi (z), & z > 0,\\
-g_\pi (-z), & z < 0 . \\
\end{array}
\right.
\label{pion_approx}
\eeq
To justify this approximation, we note that, while the exact
$t$--dependence of the non-forward gluon density in the pion is
unknown, its characteristic scale can be safely assumed to be much
harder than the $4 M_\pi^2$ involved in the extrapolation from $t = 0$
to the two--pion threshold. A reasonable guess is to suppose that the
distribution of gluons in the pion follows that of the quarks, which
would imply that the two--gluon form factor of the pion [defined in
analogy to Eq.(\ref{twogluon_def})] can be approximated by the
electromagnetic form factor,
\beq
\Gamma_\pi (t) \;\; \approx \;\; \frac{1}{1 - t/m_{2g}^2}, 
\hspace{2em}
m_{2g}^2 \; \approx \; M_{\rho}^2 \; \approx \; 0.5 \, {\rm GeV}^2
\label{twogluon_pi}
\eeq
(we assume the form factor to be independent of the gluon momentum
fraction $z$, in the region of interest). The analogy with the
nucleon, where the two--gluon form factor has been extracted from the
$t$--dependence of $J/\psi$ photoproduction data, suggests that its
scale may be even harder, $m_{2g}^2 \approx 1 \, {\rm GeV}^2$
\cite{Frankfurt:2002ka}.  Either way, one concludes that the
continuation from $t = 0$ to $4 M_\pi^2$ in the unphysical region
should have a small effect ($\sim 10\%$).

Another way of motivating the approximation Eq.(\ref{pion_approx}) is
to say that we shall assume that the intrinsic transverse size of the
pion is small compared to the transverse distances, $b$, at which we
want to study the gluon distribution in the nucleon.  This is
certainly justified as long as we are interested in the asymptotic
behavior in the strict $b \rightarrow \infty$ limit.  Later, when we
use the asymptotic expression at finite $b$ to estimate the pion cloud
contribution to the nucleon's transverse size, we shall expect
corrections from the intrinsic transverse size of the pion, see the
discussion in Section~\ref{sec_size}.

The above approximation can be stated in an alternative way, by
defining an operator in the interpolating pion field, which represents
the QCD gluon light--ray operator in the effective chiral theory. One
easily sees that Eqs.(\ref{H_g_pi}) and (\ref{pion_approx}) amount to
matching the QCD gluon operator with a pionic light--ray operator as
\be
&& n^\alpha n^\beta \; G_{\alpha}^{A\gamma} (-\lambda n/2)
\, G^A_{\gamma\beta} (\lambda n/2)
\nonumber \\
&\rightarrow & 
\; \frac{i}{2} \; n^\alpha n^\beta \; \int_{0}^1 dz \; z \, g_\pi (z) \;
\nonumber \\
&\times&  \left[ \pi^a (-z \lambda n/2) 
\stackrel{\leftrightarrow}{\partial}{}_{\!\!\alpha}
\stackrel{\leftrightarrow}{\partial}{}_{\!\!\beta}
\pi^a (z \lambda n /2) \; + \; (z \rightarrow -z) \right]
\nonumber \\
&+& \mbox{total derivatives} ,
\label{bosonized_gluon}
\ee
where $\stackrel{\leftrightarrow}{\partial} \equiv
(1/2)(\stackrel{\rightarrow}{\partial} -
\stackrel{\leftarrow}{\partial})$. We have omitted here terms
involving total derivatives contracted with the light--like vector
$n$, which contribute only for non-zero longitudinal momentum
transfer, {\it cf.}\ Eq.(\ref{nonskewed}). Computing the matrix
element of the pionic operator (\ref{bosonized_gluon}) between pion
states and comparing with the general parametrization
Eq.(\ref{H_g_pi}), one obtains $H_{g, \pi} (z, t) = g_\pi (z)$ for 
$z > 0$, in agreement with Eq.(\ref{pion_approx}). The approximation
of neglecting the $t$--dependence of the gluon distribution in the
pion is now encoded in the fact that the pionic operator does not
contain any contracted total derivatives (or, equivalently, no
transverse total derivatives), which would give rise to powers of
${\bf \Delta_\perp}^2 = -t$ in the matrix element \footnote{Note that
the usual derivative expansion for soft--pion operators is being
modified here, to the effect that {\it longitudinal} derivatives 
$(n \partial )$ are included to all orders, while {\it transverse}
derivatives are restricted to the minimum number. A similar logic was
applied in the calculation of chirally singular contributions to the
GPD's in the pion in the chiral perturbation theory approach, see
N.~Kivel and M.~V.~Polyakov, arXiv:hep-ph/0203264.}. The pionic
operator is defined also off--shell, to the extent that the above
approximations are valid. In particular, with the operator
representation of the gluon density in the pion, we are allowed to
interpret the graphs of Fig.~\ref{fig_graphs} as Feynman graphs, which
can be evaluated using standard methods.

A comment is in order concerning chiral invariance. Strictly speaking,
the QCD gluon operator should be matched with a chirally invariant
pionic operator, constructed from fields that transform according to a
non-linear realization of the chiral group, $SU(2)_L \times SU(2)_R$.
The bilinear pion operator of Eq.(\ref{bosonized_gluon}) is not
chirally invariant; however, it corresponds to the small--field limit
of such a chirally invariant operator. This simplification is
justified in our case, since we shall only evaluate the pionic
operator in the region of large impact parameters $b \gtrsim 1/M_\pi$,
where the pion field is exponentially small in $b$. Our results for
the asymptotic behavior for large $b$ below would be the same if we
started from a chirally invariant, non-linear operator.

Having at hands the representation of the QCD twist--2 gluon operator
as a pionic operator, Eq.(\ref{bosonized_gluon}), it is
straightforward to compute the pion exchange contribution to the
non-forward gluon densities in the nucleon. Substituting the pionic
representation of the gluon operator in the matrix element of
Eq.(\ref{H_g_E_g_def}), and changing the integration variable to 
$y \equiv x/z$, one obtains (here $x > 0$)
\be
H_g (x, t)_{\rm cloud}
&=& \int_{x}^1 \frac{dy}{y} \;
g_\pi (x/y) \; H_\pi (y, t) ,
\label{convolution_t_H}
\\
E_g (x, t)_{\rm cloud}
&=& \int_{x}^1 \frac{dy}{y} \;
g_\pi (x/y) \; E_\pi (y, t) .
\label{convolution_t_E}
\ee
The functions $H_\pi$ and $E_\pi$ parametrize the nucleon matrix
element of the pion light--ray operator, in complete analogy to the
gluon distributions of Eq.(\ref{H_g_E_g_def}) \footnote{In order to
arrive at the form of the pionic light--ray operator in
Eq.(\ref{H_pi_E_pi_def}) we have converted the longitudinal
derivatives $(n \stackrel{\leftrightarrow}{\partial})$ in
Eq.(\ref{bosonized_gluon}) into derivatives with respect to the
parameter $\lambda$, which can be removed by integration by parts.},
\be
&& 2 (Pn)^2 y \int_{-\infty}^\infty
\frac{d\lambda}{2\pi} \; e^{i\lambda y (Pn)} \; \langle N(P + \Delta/2) |
\nonumber \\
&\times&  \pi^a (-\lambda n/2) 
\pi^a (\lambda n /2) \; | N(P - \Delta /2) \rangle
\nonumber \\
&=& H_\pi (y, t) \; \bar u' \hat n u 
\; + \;
\frac{E_\pi (y, t)}{2 M_N} \;
\bar u' \sigma_{\mu\nu} \Delta^\mu n^\nu u .
\label{H_pi_E_pi_def}
\ee
They can be interpreted as the isoscalar non-forward ``parton densities'' 
of pions in the nucleon, with $y$ the longitudinal momentum fraction
of the pion. We emphasize that this notion is meaningful only in the 
region in which the pion virtualities are of order $M_\pi^2$;
we shall see below that this is indeed the case for $y < M_\pi / M_N$.
In the impact parameter representation [{\it cf.}\ Eq.(\ref{impact_def})], 
Eqs.(\ref{convolution_t_H}) and (\ref{convolution_t_E}) 
take the form (here $x > 0$)
\beq
g(x, b)_{\rm cloud}
\;\; = \;\; \int_{x}^1 \frac{dy}{y} \;
g_\pi (x/y) \; f_\pi (y, b) ,
\label{convolution}
\eeq
and similarly for the helicity flip--distributions.
The $b$--dependent distributions of pions in the nucleon are
defined in analogy to Eq.(\ref{impact_def}),
\beq
H_\pi (y, -{\bf \Delta}_\perp^2 ) \;\; = \;\; \int d^2 b \; 
e^{i ({\bf b} {\bf \Delta}_\perp)}
\; f_\pi (y, b) ,
\label{f_pi_N_def}
\eeq
and similarly for the helicity--flip distribution $E_{\pi} (y, t)$.
Again, these distributions are meaningful only at distances 
$b \gtrsim 1/M_\pi$, where the pion virtualities are of order
$M_\pi^2$, as will be shown below. We refer to Eq.(\ref{convolution})
as the ``pion cloud contribution'' to the gluon density in the
nucleon.  Note that the relation between the distribution of pions and
of gluons is local in impact parameter space; this is a result of our
approximation of neglecting the intrinsic transverse size of the pion.

The distribution of pions in the nucleon at impact parameters $b
\gtrsim 1/M_\pi$ is determined by chiral dynamics and can be computed
from first principles, regarding the pion and nucleon as pointlike
particles and using the phenomenological $\pi N N$ coupling. For the
moment we consider only the graphs in Fig.~\ref{fig_graphs} with
intermediate nucleon states, which dominate in the strict 
$b \rightarrow \infty$ limit; contributions with intermediate $\Delta$
states will be discussed in Section~\ref{sec_largenc} in connection
with the large--$N_c$ limit.  The essential steps of the calculation
are presented in Appendix~\ref{app_twopion}.  One computes the
$t$--dependent distribution of pions through the matrix element of
pion light--ray operator in Eq.(\ref{H_pi_E_pi_def}), and then studies
the asymptotic behavior of its Fourier transform at large $b$. This
can be done either by writing the Feynman integral in an appropriate
parameter representation (Appendix~\ref{app_feynman}), or by directly
computing the imaginary part of the $t$--dependent distribution using
the Cutkosky rules (Appendix~\ref{app_cutkosky}).  The distributions
at large $b$ are finite without cutoffs at the pion--nucleon vertices,
and independent of the type of pion--nucleon vertex (pseudoscalar or
axial vector) employed in the calculation.  The result for the
asymptotic behavior at $b \rightarrow \infty$ is
\be
f_\pi (y, b) &=&
\frac{3 g_{\pi NN}^2}{16 \pi^2} \;
\frac{y}{\bar y} \left( \frac{2 y^2 M_N^2}{\bar y} + M_\pi^2 \right) \;
\frac{e^{\displaystyle -\kappa_N b}}{\kappa_N b}
\nonumber \\[2ex]
&\equiv & f_{\pi N} (y, b) .
\label{f_pi_N_asymp}
\ee
To indicate the fact that this part of the distribution of pions is
obtained from intermediate nucleon states only, we denote it by
$f_{\pi N}$, in order to distinguish it from the contribution from
intermediate $\Delta$ states to be introduced below; a similar
notation we adopt for the $t$--dependent distributions of
Eq.(\ref{H_pi_E_pi_def}).  Here and in the following,
\beq
\bar y \;\; \equiv \;\; 1 - y. 
\eeq
At large $b$ the distribution of pions drops exponentially, with 
the ``decay constant''
\beq
\kappa_N \;\; \equiv \;\; 2 \left[ \frac{1}{\bar y}
\left( \frac{y^2 M_N^2}{\bar y} + M_\pi^2 \right) \right]^{1/2} .
\label{kappa_N}
\eeq
Note the similarity with the Yukawa potential in the $NN$ interaction.
However, in our case the decay constant depends on the longitudinal
momentum fraction of the pions in the nucleon, $y$.  This
$y$--dependence, which arises in an elementary way from the
singularities of the Feynman graphs in Fig.~\ref{fig_graphs} (see
Appendix~\ref{app_twopion}), plays a crucial role for the effects
discussed in the following.

\begin{figure}[t]
\begin{center}
\psfrag{km1ofm}{{\Large $\kappa_N^{-1} / {\rm fm}$}}
\psfrag{myy}{{\Large $y$}}
\psfrag{rrr}{{\Large $\displaystyle \frac{M_\pi}{M_N}$}}
\includegraphics[width=8.4cm,height=5.9cm]{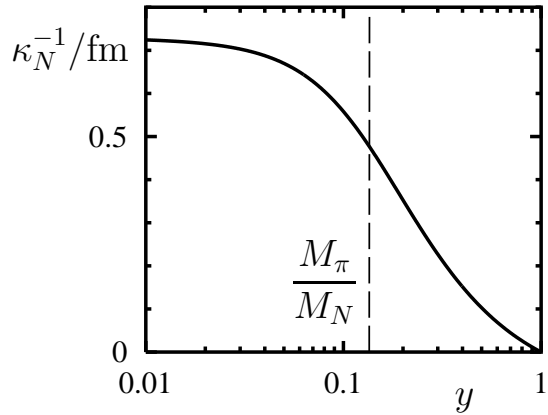}
\end{center}
\caption[]{The inverse ``decay constant'', $1/\kappa_N$, of the 
exponential tail of the distribution of pions in the nucleon
at large impact parameters, Eqs.(\ref{f_pi_N_asymp}) and (\ref{kappa_N}),
as a function of the pion momentum fraction, $y$. This quantity 
determines the transverse size of the pion cloud for given $y$. 
The dashed line marks the value $y = M_\pi / M_N = 0.14$. 
The limiting value of $1/\kappa_N$ for $y \rightarrow 0$ is
$1/(2 M_\pi) = 0.73 \, {\rm fm}$. Note that the notion of 
pion cloud contributions to the nucleon's parton distributions
is physically meaningful only for pion momentum fractions 
$y < M_\pi / M_N$, where the size of the pion cloud is 
significantly larger than those of average configurations 
in the nucleon.}
\label{fig_kappa}
\end{figure}

The inverse decay constant, $1/\kappa_N$, is a measure of the
transverse size of the distribution of pions in the nucleon. Its
dependence on $y$ exhibits two different regimes, see
Eq.(\ref{kappa_N}). For $y < M_\pi / M_N$ the size becomes independent
of the nucleon mass and approaches the value $1/(2 M_\pi )$ at 
$y = 0$. In this region the transverse size of the pion distribution
is determined by the pion mass only (it would be singular in the
chiral limit, $M_\pi \rightarrow 0$).  For $y > M_\pi / M_N$ the
transverse size is of the order of the inverse nucleon mass,
$1/M_N$. It is a slowly decreasing function of $y$, vanishing linearly
in $1 - y$ for $y \to 1$, in agreement with general
expectations. Fig.~\ref{fig_kappa} shows the transverse size
$1/\kappa_N$ as a function of $y$ obtained with the physical values of
the pion mass; the dashed line marks the value $y = M_\pi / M_N$, the
boundary between the two regimes. The above implies that the very
notion of pion cloud contributions to the nucleon's parton
distributions is meaningful only for pion momentum fractions $y <
M_\pi / M_N$, where the size of the pion cloud is significantly larger
than that of average configurations in the nucleon.

The change of the spatial size of the pion distribution with $y$ is
matched by a corresponding change of the pion virtualities.  From
Eqs.(\ref{H_pi_N_kperp}), (\ref{s_pm}) and (\ref{s_min_N_app}) in
Appendix~\ref{app_twopion} one sees that the minimum value of the
virtuality (taken with positive sign) of the pions in the distribution
Eq.(\ref{H_pi_E_pi_def}), {\it i.e.}, the virtuality of the pion lines
in the graphs of Fig.~\ref{fig_graphs}, is
\beq
s_{{\rm min}, N} \;\; = \;\; \frac{y^2 M_N^2}{\bar y} ,
\label{s_min_N}
\eeq
see also the discussion in Ref.\cite{Koepf:1995yh}, where this
quantity is denoted by $-t_{\rm min}$. For $y$ of order unity the pion
virtuality is of the order $M_N^2$, and the notion of pion exchange
becomes meaningless. It is only for values $y < M_\pi / M_N$ that the
pion virtuality reduces to values of the order $M_\pi^2$, where the
pion can be regarded as soft.  As we see from the above discussion,
this is the region where the transverse size of the pion distribution,
$1/\kappa_N$, is of the order $1/M_\pi$.

\begin{figure}[t]
\begin{center}
\psfrag{xlarger}{{\large $x > M_\pi / M_N$}}
\psfrag{xsmaller}{{\large $x \ll M_\pi / M_N$}}
\psfrag{oomn}{\large $1/M_N$}
\psfrag{oomp}{\large $1/M_\pi$}
\includegraphics[width=8.3cm,height=6.7cm]{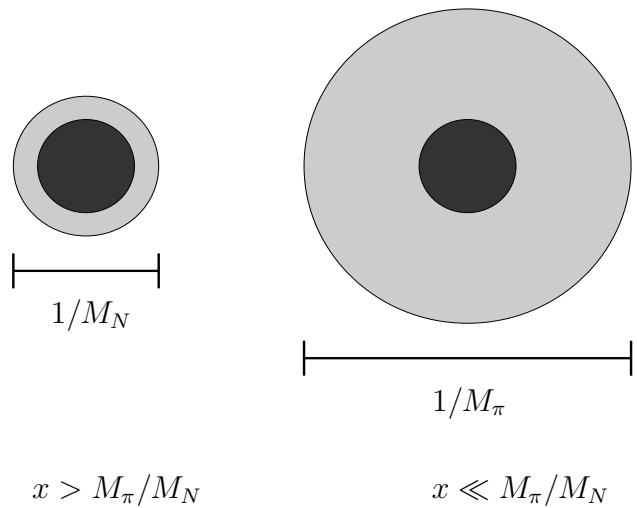}
\end{center}
\caption[]{Visualization of the characteristic transverse size of the
pion cloud contribution to the gluon density in the nucleon, as
defined by Eqs.(\ref{convolution}), (\ref{f_pi_N_asymp}) and
(\ref{kappa_N}). $x > M_\pi / M_N$: The transverse size of the pion
cloud contribution (light shaded disc) is of order $1/M_N$, comparable
to that of average configurations in the nucleon wave function
producing the ``bulk'' of the gluon distribution (dark shaded
disc). In this regime the notion of pion cloud contributions is not
meaningful. $x \ll M_\pi / M_N$: The size of the pion cloud
contribution grows to $1/M_\pi$. Now the pion cloud contribution is
distinct from the ``bulk'' of the gluon distribution.  The result in
an increase of the overall transverse size of the nucleon (see
Section~\ref{sec_size}).}
\label{fig_twoscale}
\end{figure}
The transverse size of the distribution of pions in the nucleon
determines that of the pion cloud contribution to the gluon density,
Eq.(\ref{convolution}). The convolution integral in
Eq.(\ref{convolution}) runs over values $y > x$. For $x > M_\pi / M_N$
one integrates only over the region where the size of the distribution
of pions in the nucleon is $\sim 1/M_N$, and the size of the pion
cloud contribution to the gluon density will be accordingly. 
For $x < M_\pi / M_N$, however, the integral extends also into the
region $y < M_\pi / M_N$ where the size of the distribution of pions
is $\sim 1/M_\pi$, so the size of the pion cloud contribution to the
gluon density grows to $\sim 1/M_\pi$. Note that for the gluon
distribution at distances $b \sim 1/M_\pi$ to become sizable one
really needs $x$ significantly smaller than $M_\pi / M_N$, so that in
addition to $y$ also the momentum fraction of the gluon in the pion,
$z = x/y$, can become small.  To summarize, the qualitative change in
the transverse size of the distribution of pions between 
$y > M_\pi / M_N$ and $y < M_\pi / M_N$ ({\it cf.}\
Fig.~\ref{fig_kappa}) directly translates into a corresponding change
of the size of the pion cloud contribution to the gluon density in the
nucleon. This behavior is schematically illustrated in
Fig.~\ref{fig_twoscale}, where the transverse size of the gluon
density is represented by a disc in the transverse plane.  The dark
shaded disc indicates contributions from average configurations in the
nucleon wave function to the gluon density, whose transverse size is
of order $1/M_N$. The light shaded disc indicates the pion cloud
contributions, as defined by Eq.(\ref{convolution}). 
For $x > M_\pi / M_N$ the pion cloud contributions are ``hidden'' in
the average configurations.  In this region of $x$ the asymptotic
expression (\ref{convolution}) is only of symbolic significance, since
the chiral contributions to the gluon density are not special in any
sense. It is only for $x \ll M_\pi / M_N$ that the pions are free to
propagate over transverse distances of order $1/M_\pi$, and the pion
cloud contributions become a distinct feature of the gluon density in
the nucleon. As we have seen above, this is also the region where the
pion virtuality is of order $M_\pi^2$, and our approximations are
justified.  The existence of these two different regimes has important
consequences for the overall average transverse size squared of the
nucleon, as will be shown in Section~\ref{sec_size}.

To conclude this discussion, we return to the question of the role of
the intrinsic transverse size of the pion. The $t$--dependence of the
non-forward gluon density in the pion given by Eq.(\ref{twogluon_pi})
(assumed to be independent of the gluon momentum fraction, $z$, in the
region of interest) would imply an exponential decay of the impact
parameter--dependent gluon distribution in the pion at large impact
parameters (with respect to the center of pion) of the form
\beq
g_\pi (y, b') \;\; \sim \;\; \frac{m_{2g}^2}{\sqrt{32 \pi}} 
\frac{e^{\displaystyle -m_{2g} b'}}{\sqrt{m_{2g} b'}} .
\eeq
The characteristic size of this distribution, 
$m_{2g}^{-1} \approx 0.2 \ldots 0.3\, {\rm fm}$, is smaller than 
the size of the distribution of pions in the nucleon in the region 
$y \ll M_\pi / M_N$, $1/(2 M_\pi) = 0.73 \, {\rm fm}$, 
by a factor of $\sim 3$. This ratio gives an estimate of the accuracy 
of our ``two--scale picture'', {\it cf.}\ Fig.~\ref{fig_twoscale}.
Note that when computing the average transverse size squared
of the nucleon in Section~\ref{sec_size} the relevant ratio is
that of the transverse {\it areas} covered by the distributions,
so we can expect a reasonable accuracy in this case.
\section{Distribution of pions in the large--$N_c$ limit}
\label{sec_largenc}
So far we have studied the asymptotic behavior of the gluon
distribution at large impact parameters invoking the phenomenological
notion of soft pion exchange. It is interesting to investigate the
large--$b$ behavior also in a different but related approach to chiral
dynamics, namely the large--$N_c$ limit of QCD, where the nucleon is
described as a classical soliton of the effective chiral theory. It
turns out that in this approach one can relate the distribution of
pions in the nucleon at large $b$ directly to the large--distance
``tail'' of the classical pion field in the soliton rest frame, which
allows for a simple geometric interpretation. The functional form and
the magnitude of this ``tail'' are completely determined by chiral
symmetry, and thus independent of the particular dynamical model used
to generate stable soliton solutions. This allows us to conduct the
following discussion at a general level, assuming only a ``generic''
chiral soliton picture of the nucleon.
\par
We begin by studying the $t$--dependent distribution of pions in the
nucleon, Eq.(\ref{H_pi_E_pi_def}), in the large--$N_c$ limit.
Following the standard procedure for form factors, the $1/N_c$
expansion is performed in the Breit frame, in which the three--momenta
of the incoming and outgoing nucleons are $O(1)$ in $N_c$, and thus
suppressed compared to the energies, which are $O(N_c)$ [the nucleon
mass, $M_N$, is $O(N_c)$].  For the average momentum $P$ and the
momentum transfer $\Delta$ [{\it cf.}\ Eq.(\ref{H_g_E_g_def})] this
implies
\be
P^0 &=& M_N \; + \; O(1) ,
\\
P^i &=& 0, 
\\
\Delta^0 &=& 0, 
\\
\Delta^i &=& O(1) .
\ee 
Choosing the spatial component of the light--like vector $n$ along the
3--direction, $n^\mu = (1,0,0,-1)$, the condition that $(n\Delta) = 0$
requires $\Delta^3 = 0$, {\it i.e.}, 
${\bf\Delta}_\perp^2 = -t = O(1)$.  Since the impact parameter 
${\bf b}$ is Fourier--conjugate to ${\bf \Delta}_\perp$, this implies
that we are studying the transverse spatial distributions at distances
$b = O(1)$ in $1/N_c$. This is natural in view of the fact that
typical nucleon radii (electromagnetic, axial) are $O(1)$ in the
large--$N_c$ limit. Concerning the pion longitudinal momentum
fraction, since $M_\pi / M_N = 1/N_c$ we suppose that we compute the
distribution of pions in the nucleon for ``average'' values of the
pion momentum fraction, $y \sim 1/N_c$.

\begin{figure}[t]
\begin{center}
\psfrag{lambda}{{\Large $\lambda$}}
\psfrag{bb}{{\Large $b$}}
\includegraphics[width=8.4cm,height=5.6cm]{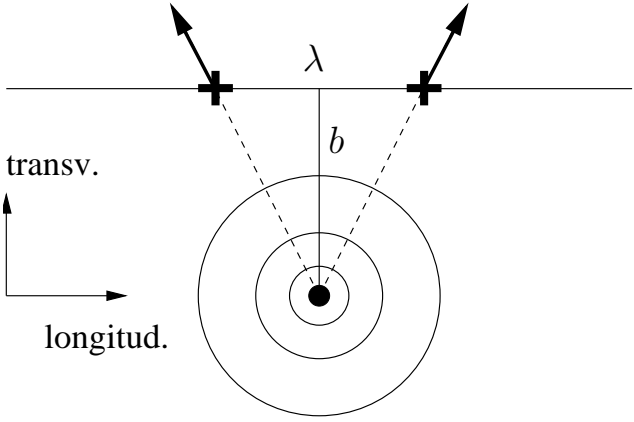}
\end{center}
\caption[]{Schematic illustration of the calculation of the
distribution of pions in the nucleon at large impact parameters in the
large--$N_c$ limit, $f_\pi (y, b)_{\mbox{\scriptsize large--$N_c$}}$,
Eq.(\ref{f_pi_largenc_int}). Shown is a 2--dimensional projection of
3--dimensional coordinate space in the soliton rest frame, containing
the longitudinal and one transverse direction (say, the 3--1
plane). The classical pion field of the soliton is indicated by the
circles; the direction of the field in isospin space at a given point
is determined by the radius vector.  The positions ${\bf r}_-$ and
${\bf r}_+$ at which the pion field is evaluated are indicated by
crosses. Their separation in the longitudinal direction is
$\lambda$. When integrating over the longitudinal coordinates
$\lambda$ and $\omega$, the points move along the line indicated in
the figure, which passes the soliton at a transverse distance $b$ from
the center.}
\label{fig_soliton}
\end{figure}

It is then straightforward to compute the matrix element in
Eq.(\ref{H_pi_E_pi_def}) in leading order of the $1/N_c$ expansion in
a ``generic'' chiral soliton picture of the nucleon.  One evaluates
the pionic operator in the classical pion field of the soliton and
projects on nucleon states of definite spin/isospin and momentum by
integrating over collective (iso--) rotations and translations of the
classical soliton with appropriate collective wave functions
\cite{Adkins:1983ya}. Taking the Fourier transform with respect to the
momentum transfer ${\bf \Delta}_\perp$ we obtain a simple expression
for the isoscalar distribution of pions at large impact parameters,
\be
f_\pi (y, b)_{\mbox{\scriptsize large--$N_c$}}
&=& 2 M_N^2 y \int_{-\infty}^\infty\frac{d\lambda}{2\pi} \;
e^{i\lambda y M_N} \; \int_{\infty}^\infty d\omega
\nonumber \\
&\times& 
\pi^a_{\rm asymp} ({\bf r}_- ) \pi^a_{\rm asymp} ({\bf r}_+) ,
\label{f_pi_largenc_int}
\ee
where
\beq
{\bf r}_\mp \;\; \equiv \;\; {\bf b} 
+ \left( \omega \mp \frac{\lambda}{2} \right) {\bf e}_3 
\eeq
are three--dimensional coordinate vectors in the frame where the 
soliton is at rest and centered at the origin, and
\beq
\pi^a_{\rm asymp} ({\bf r}) \;\; \equiv \;\; 
\frac{3\, g_{\pi NN}}{8 \pi M_N} \left( \frac{1}{r^2} + \frac{M_\pi}{r}
\right) e^{\displaystyle -M_\pi r} \; \frac{r^a}{r}
\eeq
is the ``Yukawa tail'' of the classical soliton at large distances,
which is obtained from solving the linearized field equations for the
pion field, see {\it e.g.}\ Ref.\cite{Zahed:1986qz}. A visual
representation of the integral (\ref{f_pi_largenc_int}) defining the
distribution of pions is given in Fig.~\ref{fig_soliton}. At large
$b$, which is where Eq.(\ref{f_pi_largenc_int}) applies, the integrals
over the longitudinal coordinates $\lambda$ and $\omega$ can be
computed in the saddle point approximation. For the leading asymptotic
behavior we find
\be
f_\pi (y, b)_{\mbox{\scriptsize large--$N_c$}}
&=& \frac{9\, g_{\pi NN}^2}{16 \pi^2}
\; y (2 y^2 M_N^2 + M_\pi^2) \; 
\nonumber \\
&\times&
\frac{e^{\displaystyle -\kappa_\infty b}}{\kappa_\infty b} ,
\label{f_pi_largenc_asymp}
\ee
where
\beq
\kappa_\infty \;\; \equiv \;\; 2 \left( y^2 M_N^2 + M_\pi^2 \right)^{1/2} .
\label{kappa_largenc}
\eeq
%
%

When comparing the large--$N_c$ result, Eqs.(\ref{f_pi_largenc_asymp})
and (\ref{kappa_largenc}), with the result of the phenomenological
soft--pion exchange graphs, Eqs.(\ref{f_pi_N_asymp}) and
(\ref{kappa_N}), we need to take into account that the large--$N_c$
result implies $y \sim 1/N_c \ll 1$. Indeed, we see that for $y \ll 1$
one has $\kappa_\infty = \kappa_N$, and the functional forms of
Eqs.(\ref{f_pi_N_asymp}) and (\ref{f_pi_largenc_asymp})
coincide. However, we observe that the coefficient of the large--$N_c$
result is larger than that of the pion loop calculation by a factor of
3:
\beq
f_\pi (y, b)_{\mbox{\scriptsize large--$N_c$}}
\; = \; 3 \, f_{\pi N} (y, b)
\hspace{2em} (y \; \sim \; 1/N_c ).
\label{paradox}
\eeq
This paradox is resolved by noting that in the large--$N_c$ limit the 
nucleon and the Delta resonance are degenerate,
$M_\Delta - M_N \sim 1/N_c$. Thus, the pion loop calculation in this
case should include $\Delta$ intermediate states on the same footing as 
nucleon ones, {\it i.e.}, the relevant distribution of pions is now
\beq
f_\pi (y, b) \;\; = \;\; f_{\pi N} (y, b) \; + \; 
f_{\pi\Delta} (y, b) ,
\label{f_pi_total}
\eeq
where $f_{\pi\Delta} (y, b)$ denotes the contribution from the
graphs in Fig.~\ref{fig_graphs} with $\Delta$ intermediate states.

To verify the equivalence of the two approaches explicitly, we compute
the $\Delta$ contribution to the distribution of pions at large $b$ 
in the approach of Section~\ref{sec_large}, describing the $\Delta$ 
resonance by an elementary Rarita--Schwinger field, and assuming
a phenomenological $\pi N \Delta$ vertex. Details are presented in 
Appendix~\ref{app_Delta}. In the general case, $M_\Delta > M_N$ 
and $0 < y < 1$ arbitrary, we obtain for the leading large--$b$ behavior 
\be
f_{\pi\Delta} (y, b) &=&
\frac{g_{\pi N\Delta}^2}{12 \pi^2 \, M_N^2} \; \frac{y}{\bar y^2} \;
\left[ \frac{M_\Delta^2}{\bar y} + \bar y M_N^2 
\phantom{\frac{0}{0}} \right.
\nonumber \\
&& \left. 
- \frac{1}{2} (M_N - M_\Delta)^2 + \frac{M_\pi^2}{2} \right]
\nonumber \\
&\times& \left[ \frac{2 y^2 M_N^2}{\bar y} 
+ (1 + y) M_\pi^2 + \frac{2 y}{\bar y} (M_\Delta^2 - M_N^2)
\right.
\nonumber \\ 
&& \left. + \frac{\bar y}{4 M_\Delta^2} 
(M_\Delta^2 - M_N^2 + M_\pi^2)^2 \right]
\nonumber \\ 
&\times& \frac{e^{\displaystyle -\kappa_\Delta b}}{\kappa_\Delta b} ,
\label{f_pi_Delta_asymp}
\ee
where 
\beq
\kappa_\Delta 
\;\; \equiv \;\; 2 \left[ \frac{1}{\bar y} \left( \frac{y^2 M_N^2}{\bar y} 
+ \frac{y (M_\Delta^2 - M_N^2 )}{\bar y}  + M_\pi^2 \right) \right]^{1/2} .
\label{kappa_Delta}
\eeq
Note that for $M_\Delta > M_N$ the exponential decay
of the $\Delta$ contribution is faster than that of the $N$ 
contribution, {\it cf.}\ Eqs.(\ref{f_pi_N_asymp}) and (\ref{kappa_N}). 
In the large--$N_c$ limit, however, where $M_N, M_\Delta \sim N_c, 
\; M_N - M_\Delta \sim 1/N_c$, and $M_\pi \sim 1$, and for 
$y \sim 1/N_c$, we can neglect the
$y (M_\Delta^2 - M_N^2) / \bar y$ term in Eq.(\ref{kappa_Delta}),
and thus have $\kappa_\Delta = \kappa_N + O(1/N_c)$, {\it i.e.}, the two 
contributions exhibit the same exponential decay. 
Using the same relations we can simplify also the pre-exponential factors. 
Finally, taking into account that in the large--$N_c$ 
limit \cite{Adkins:1983ya}
\beq
g_{\pi N\Delta} \;\; = \;\; \frac{3}{2} g_{\pi N N} ,
\eeq
we see that the $\Delta$ contribution in this limit is exactly 
twice as large as the $N$ contribution, Eq.(\ref{f_pi_N_asymp}):
\beq
f_{\pi\Delta} (y, b)
\; = \; 2 \, f_{\pi N} (y, b)
\hspace{2em} \mbox{(large--$N_c$).}
\label{pi_Delta_pi_N_largenc}
\eeq
The sum of the $N$ and $\Delta$ contributions, Eq.(\ref{f_pi_total}), 
in the large--$N_c$ limit is thus exactly 3 times the $N$ contribution
alone, and coincides with the result of the calculation using the 
classical soliton field, Eq.(\ref{f_pi_largenc_asymp}).

To summarize, in the large--$N_c$ limit the calculation of the
distribution of pions from the ``tail'' of the classical soliton and
from phenomenological soft--pion exchange graphs, including both $N$
and $\Delta$ intermediate states, give the same result.  Due to the
degeneracy of the $N$ and $\Delta$ the large--$N_c$ limit leads to a
different asymptotic behavior at large $b$ compared to an ``exact''
phenomenological approach based on physical states: In the ``exact''
approach the $\Delta$ contributions would be exponentially suppressed
relative to the $N$ ones and thus formally subleading, while in the
large--$N_c$ limit both come with the same exponential factor and
contribute on the same footing, with the $\Delta$ contribution even
dominating due to the larger spin--isospin degeneracy. The fact that
the $\Delta$ contribution is sizable in the large--$N_c$ limit
suggests that it is quantitatively important also in the ``exact''
approach and should be included, even though it is formally
subleading.  We shall indeed include it in our estimate of the
transverse size of the nucleon in Section~\ref{sec_size}.

A comment is in order concerning the overall order of the pion cloud
contribution to the gluon density within the $1/N_c$ expansion. 
If one infers the $N_c$--scaling of $g_{\pi NN}$ from the 
Goldberger--Treiman relation, $g_{\pi NN} = M_N g_A/F_\pi$,
in which the isovector axial coupling of the nucleon, $g_A$,
scales as $N_c$, and the pion decay constant, $F_\pi$, 
as $N_c^{1/2}$, one is led to $g_{\pi NN} \sim N_c^{3/2}$
\footnote{It is well--known that the $N_c$--scaling of $g_{\pi NN}$
derived from the Goldberger--Treiman relation is at variance with the
standard $N_c$--scaling assumed for meson--baryon couplings, 
$g \sim N_c^{1/2}$. For a discussion of this problem, see
{\it e.g.}\ H.~M\"uther, C.~A.~Engelbrecht and G.~E.~Brown,
Nucl.\ Phys.\ A {\bf 462}, 701 (1987).}.
This would mean that the distribution of pions in the nucleon at
large $b$, Eq.(\ref{f_pi_largenc_asymp}), as a function of $y$, 
behaves as 
\beq
f_\pi (y, b)_{\mbox{\scriptsize large--$N_c$}}
\;\; \sim \;\; N_c^2 \times \mbox{function} (N_c y, \, b) .
\label{f_pi_scaling}
\eeq
This scaling behavior is analogous to that of the gluon distribution 
in the nucleon in the large--$N_c$ limit \cite{Diakonov:1996sr}.
Assuming that the gluon distribution in the pion, $g_\pi (z)$, 
does not scale explicitly with $N_c$, Eq.(\ref{f_pi_scaling}) 
would imply that the large--$N_c$ scaling of the large--$b$ 
contribution to the gluon density, as defined by 
the convolution integral (\ref{convolution}), is of the form 
\beq
g(x, b)_{\rm cloud} \;\; \sim \;\;
N_c^2 \times \mbox{function} (N_c x, \, b),
\label{g_scaling}
\eeq 
{\it i.e.}, the same as for the gluon distribution at average values
of $b$. This seems natural, as going to large $b$ should not change
the $N_c$--scaling behavior of the gluon distribution [remember that
we consider $b = O(1)$ in $1/N_c$].

Our discussion of pion cloud contributions to the nucleon's parton
distributions in the large--$N_c$ limit so far pertained to isoscalar
distributions --- specifically, the gluon distribution.  To complete
the picture, it is instructive to consider within the same approach
also isovector distributions, which are known to be suppressed
relative to the isoscalar ones in the $1/N_c$ expansion (for the
polarized distributions, the relative order would be reversed)
\cite{Diakonov:1996sr}.  The simplest example is actually the flavor
asymmetry of the sea quark distributions in the proton, 
$\bar q_3 (x) \equiv \bar d(x) - \bar u(x)$.  This distribution has
extensively been studied within the traditional pion cloud model, in
which no restriction is placed on impact parameters, but the the pion
loop integrals are cut off by form factors at the $\pi NN$ and 
$\pi N \Delta$ vertices, see Refs.\cite{Kumano:1997cy} for a
review. We can, of course, also study it in our approach, where pion
exchange is considered only at impact parameters $b \sim 1/M_\pi$.
The analogue of Eq.(\ref{convolution}), describing the large--$b$
asymptotics of the flavor asymmetry, including both $N$ and $\Delta$
contributions, is (see {\it e.g.}\ Ref.\cite{Koepf:1995yh})
\be
\bar q_3 (x, b) &=& \int_x^1 \frac{dy}{y} \; 
\bar q_\pi (x/y) 
\nonumber \\
&\times& \left[ \frac{2}{3} \; f_{\pi N} (y, b) \; - \; 
\frac{1}{3} \; f_{\pi\Delta} (y, b) \right] .
\label{asym_convolution}
\ee
Here $\bar q_\pi (z) \equiv \bar d_{\pi+} (z) = \bar u_{\pi-} (z)$ is
the antiquark distribution in the charged pions, which may safely be
approximated by the valence quark distribution, 
$\bar q_\pi (z) \approx (1/2) v_\pi (z)$ in the conventions of
Ref.\cite{Gluck:1999xe}. The distributions of pions in the nucleon,
$f_{\pi N}$ and $f_{\pi\Delta}$ are the isoscalar distributions (sum
of the distributions of $\pi^+, \pi^-$ and $\pi^0$), as defined in
Eqs.(\ref{f_pi_N_asymp}) and (\ref{f_pi_Delta_asymp}); the isovector
nature of $\bar q_3 (x, b)$ is accounted for by the factors $2/3$ and
$-1/3$.  Note that here the $\Delta$ contribution enters with opposite
sign relative to the $N$ one. In the large--$N_c$ limit
$f_{\pi\Delta}$ is exactly twice as large as $f_{\pi N}$, see
Eq.(\ref{pi_Delta_pi_N_largenc}), and the two contributions in
Eq.(\ref{asym_convolution}) cancel exactly. This is as it should be:
On general grounds, the isovector unpolarized parton distributions are
suppressed at large $N_c$ compared to the isoscalar ones
\cite{Diakonov:1996sr}, {\it i.e.}, they scale as 
$N_c \times \mbox{function} (N_c x)$ as opposed to
Eq.(\ref{g_scaling}), and if the leading pion cloud contribution did
not cancel it would, following the logic of the preceding paragraph,
induce a contribution with the same $N_c$--scaling as in the isoscalar
case, in violation of the general scaling behavior. Thus, we see that,
also in the isovector case, the large--$b$ contributions due to pion
exchange respect the basic $N_c$--scaling properties of the parton
distributions.
\section{Transverse size of the nucleon at small $x$}
\label{sec_size}
We now proceed to investigate the relevance of the asymptotic behavior
of the impact parameter--dependent distribution at large $b$, derived
in Sections~\ref{sec_large} and \ref{sec_largenc}, for the
$x$--dependence of the two--gluon form factor of the nucleon. For
simplicity, we concentrate on the overall transverse size of the
nucleon, $\langle b^2 \rangle$, which determines the $t$--slope of the
two--gluon form factor at small $t$.  This will reveal an interesting
phenomenon --- an increase of the transverse size of the nucleon at
small $x$ due to pion cloud contributions.

The average impact parameter squared is defined as an integral over
all values of $b$, {\it cf.}\ Eq.(\ref{b2_def}). When computing this
quantity we have to consider the large--$b$ contributions, where the
gluon distribution is described by Eqs.(\ref{convolution}) and
(\ref{kappa_N}), together with contributions from average
configurations in the nucleon, which account for the bulk of the gluon
distribution.  The size of these configurations is determined by the
binding of the valence quarks in the nucleon. In a schematic
``two--component'' picture we can write
\be
\langle b^2 \rangle \!
&=& \! \frac{\displaystyle \int \!\! d^2b \, b^2 \; 
\left[ g(x, b)_{\rm bulk} + \Theta (b > b_0) \,
g(x, b)_{\rm cloud} \right]}{\displaystyle g(x)}
\nonumber \\
&\equiv& \langle b^2 \rangle_{\rm bulk} 
\; + \; \langle b^2 \rangle_{\rm cloud} .
\label{core_cloud}
\ee 
The $b^2$--integral in the numerator is computed in two separate
pieces, while in the denominator we have the total gluon distribution
(bulk plus cloud) for the given value of $x$. The integral over the
pion cloud contribution is restricted to values $b > b_0$; the cutoff
$b_0$ will be specified below.  To estimate the ``bulk'' contribution
to $\langle b^2 \rangle$ we should look to the slope of the elastic
``two--gluon form factor'' of the nucleon, {\it cf.}\
Eqs.(\ref{twogluon_def}) and (\ref{b2_from_slope}).  Unfortunately,
this form factor cannot be measured directly in experiment.  For a
rough estimate we can take the form factor of a quark operator which
does not receive contributions from the pion cloud, {\it e.g.}\ the
axial form factor of the nucleon. This gives an estimate of the
``bulk'' contribution to the transverse size of
\beq
\langle b^2 \rangle_{\rm bulk} 
\;\; \approx \;\; \frac{2}{3} \, \langle r^2 \rangle_{\rm axial} ,
\label{bulk_axial}
\eeq
where the factor $2/3$ arises from converting the ``three-dimensional'' 
axial charge radius into the ``two-di\-men\-sional'' average of $b^2$,
{\it cf.}\ Eq.(\ref{b2_from_slope}) and after. With the experimental value 
$\langle r^2 \rangle_{\rm axial} = 0.46 \, {\rm fm}^2$ this comes to 
\beq
\langle b^2 \rangle_{\rm bulk} \;\; \approx \;\; 0.3 \, {\rm fm}^2 .
\label{bulk_num}
\eeq
This simple estimate reproduces well the measured $t$--slope of the
$J/\psi$ photoproduction cross section at fixed--target energies
\cite{Frankfurt:2002ka}.

Consider now the $x$--dependence of the ``cloud'' contributions to
$\langle b^2 \rangle$, which is computed by integrating the asymptotic
expression (\ref{convolution}) over $b$.  We have seen in
Section~\ref{sec_large} that the characteristic transverse size of the
pion cloud contribution to the gluon density is of order $1/M_N$ for
$x > M_\pi / M_N$, and becomes of order $1/M_\pi$ for 
$x \ll M_\pi / M_N$, see Fig.~\ref{fig_twoscale}. We thus expect an
increase in $\langle b^2 \rangle_{\rm cloud}$ at small $x$, which
should set in at approximately $x \sim M_\pi / M_N$.  Assuming that
the bulk contribution $\langle b^2 \rangle_{\rm bulk}$ does not change
much over the region of $x$ considered, this would result in an
increase in the overall transverse size of the nucleon, 
$\langle b^2 \rangle$, see Eq.~(\ref{core_cloud}).

%
%
\begin{figure}[t]
\begin{center}
\psfrag{xgxxxx}{{\Large $x g(x)$}}
\psfrag{myx}{{\Large $x$}}
\includegraphics[width=8.4cm,height=5.9cm]{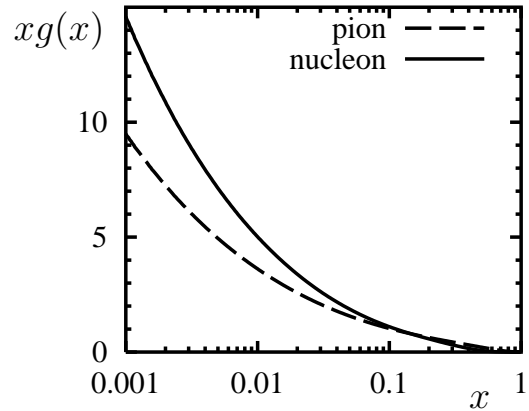}
\end{center}
\caption[]{The gluon distribution $x g(x)$ in the pion (dashed line)
and the nucleon (solid line) at $Q^2 = 4\, {\rm GeV}^2$, according to
the parametrizations of Refs.\cite{Gluck:1994uf,Gluck:1999xe}.}
\label{fig_gdist}
\end{figure}
To investigate this effect numerically, we choose the GRV
parametrization of the LO gluon distribution in the nucleon at a scale
$Q^2 = 4\, {\rm GeV}^2$ \cite{Gluck:1994uf}, and the parametrization
of Ref.\cite{Gluck:1999xe} for the gluon distribution in the pion. As
shown in Fig.~\ref{fig_gdist}, the rise of the two distributions at
small $x$ is similar; the ratio of the pion and nucleon distributions
for $x < 0.01$ is approximately $2/3$, as would be expected of gluons
generated radiatively from valence quarks at a low scale.

\begin{figure}[t]
\begin{center}
\psfrag{mN}{{\large $N$}}
\psfrag{mD}{{\large $\Delta$}}
\psfrag{intfybb2}{\parbox[r]{5em}{\large $\displaystyle \int db \, b^2$ 
\\[0.3ex] $\times f_{\pi} (y, b)$}}
\psfrag{myy}{{\Large $y$}}
\includegraphics[width=8.4cm,height=5.9cm]{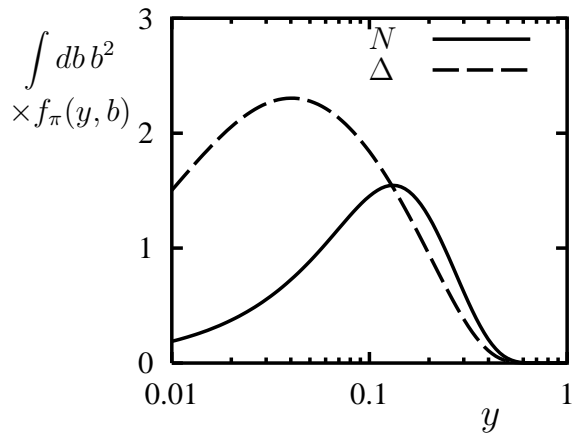}
\end{center}
\caption[]{The $b^2$--weighted integral of the distributions of pions 
in the nucleon, Eq.(\ref{f_pi_b2_int}), with a lower cutoff 
$b_0 = \langle b^2 \rangle_{\rm bulk}^{1/2} = 0.55\, {\rm fm}$, 
as a function of the pion momentum fraction, $y$.
The values of the integrals are given in units of $b_0^2$.
Shown are the contributions from $N$ (solid line) and $\Delta$ 
intermediate states. These integrals enter in the calculation of the 
pion cloud contribution to the transverse size of the nucleon, 
$\langle b^2 \rangle_{\rm cloud}$, {\it cf.}\ Eq.(\ref{core_cloud}).}
\label{fig_fpib2}
\end{figure}

Following the discussion in Section~\ref{sec_largenc}, we include in
the pion cloud contribution to the transverse size also contributions
from $\Delta$ intermediate states, which, although theoretically
subleading at large $b$, are expected to be important
quantitatively. We define the total distribution of pions in the
nucleon as in Eq.(\ref{f_pi_total}), and evaluate $f_{\pi N}$ and
$f_{\pi\Delta}$ using the asymptotic expressions at large $b$,
Eqs.(\ref{f_pi_N_asymp}) and (\ref{f_pi_Delta_asymp}). We use the
physical values for the $N$ and $\Delta$ masses; for the coupling
constants we take $g_{\pi NN} = 13.5$ and 
$g_{\pi N\Delta} = (3/2)\, g_{\pi NN}$, which is close to the
phenomenological value \cite{Machleidt:hj}.  We then compute the
$b^2$--weighted integral of the distributions with a lower cutoff
$b_0$, {\it cf.}\ Eq.(\ref{core_cloud}), whose value we take to be the
size of the bulk of the gluon distribution in the nucleon, $b_0 =
\langle b^2 \rangle_{\rm bulk}^{1/2} = 0.55 \, {\rm fm}$.  
The integrated distributions
\beq
\int\limits_{b > b_0} d^2 b \; b^2 \; f_{\pi N, \pi\Delta} (y, b) ,
\label{f_pi_b2_int}
\eeq
are shown in Fig.~\ref{fig_fpib2} (in units of $b_0^2$), as functions
of $y$.  One sees that the maximum of the $N$ contribution is
approximately at $y \sim M_\pi / M_N$. The $\Delta$ contribution sets
in at smaller values of $y$ compared to $N$ --- a consequence of the
different $y$--dependence of the exponential decay constants,
Eqs.(\ref{kappa_N}) and (\ref{kappa_Delta}) --- but eventually becomes
larger than the $N$ contribution at smaller $y$, in qualitative
agreement with the large--$N_c$ relation,
Eq.(\ref{pi_Delta_pi_N_largenc}).

\begin{figure}[t]
\begin{center}
\psfrag{piNonly}{{\large $N$ only}}
\psfrag{piNppiD}{{\large $N + \Delta$}}
\psfrag{b2-cloud}{\parbox[r]{5em}{\large $\langle b^2 \rangle_{\rm cloud}$ 
\\[.3ex] \large (${\rm fm^2}$)}}
\psfrag{myx}{\large $x$}
\includegraphics[width=8.4cm,height=5.9cm]{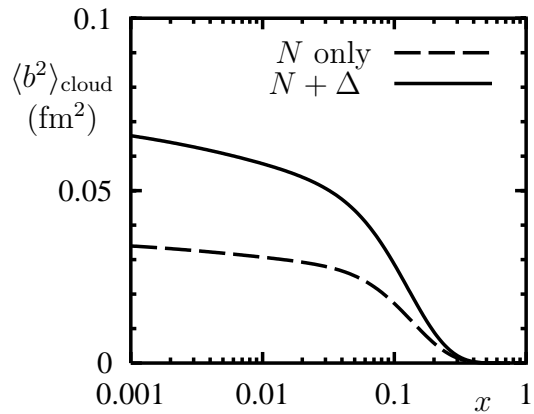}
\end{center}
\caption[]{The pion cloud contribution to the average 
transverse size squared of the gluon distribution in the nucleon, 
$\langle b^2 \rangle_{\rm cloud}$, Eq.(\ref{core_cloud}), 
as a function of $x$. The values shown here
were obtained by integrating over the large--$b$ tail of the gluon 
distribution, Eq.(\ref{convolution}),
with a lower cutoff $b_0 = \langle b^2 \rangle_{\rm bulk}^{1/2}
= 0.55\, {\rm fm}$. {\it Dashed line:} Contribution from 
$N$ intermediates states, {\it cf.}\ Eq.(\ref{f_pi_N_asymp}).
{\it Solid line:} Sum of $N$ and $\Delta$ contributions, 
{\it cf.}\ Eq.(\ref{f_pi_Delta_asymp}). These results should be 
compared with the ``bulk'' contribution to the nucleon size, 
$\langle b^2 \rangle_{\rm bulk} = 0.3 \, {\rm fm}^2$.}
\label{fig_conv}
\end{figure}

The results for $\langle b^2 \rangle_{\rm cloud}$ are presented in
Fig.~\ref{fig_conv}. The dashed line shows the result for $\langle b^2
\rangle_{\rm cloud}$ obtained from $N$ intermediate states only, 
{\it cf.}\ Eq.(\ref{f_pi_N_asymp}).  The solid line shows the sum of
the contributions from $N$ and $\Delta$ intermediate states (with
physical $N$ and $\Delta$ masses), {\it cf.}\
Eq.(\ref{f_pi_Delta_asymp}).  One sees that the onset of the $\Delta$
contribution is ``postponed'' to smaller values of $x$ compared to the
$N$ one, and that at $x \sim 10^{-2}$ the $\Delta$ contribution makes
about half of the total result \footnote{The relative importance of
$N$ and $\Delta$ contributions here is similar to what was obtained
for the flavor--singlet singlet sea quark distributions in the meson
cloud model with soft pion--nucleon form factors in
Ref.\cite{Koepf:1995yh}.}.  These results for 
$\langle b^2 \rangle_{\rm cloud}$ shown in Fig.~\ref{fig_conv} should
be compared with the bulk contribution to the nucleon size, 
$\langle b^2 \rangle_{\rm bulk}$, which we estimated at 
$\sim 0.3 \, {\rm fm^2}$, {\it cf.}  Eqs.(\ref{bulk_axial}) and
(\ref{bulk_num}). We see that, with our estimates of the parameters,
the increase in the total transverse size of the nucleon due to pion
cloud contributions for $x \ll M_\pi / M_N$ amounts to $\sim 20\%$ of
the total transverse size of the nucleon.  This is in qualitative
agreement with the analysis of Ref.~\cite{Frankfurt:2002ka}, which
concluded that the nucleon two--gluon form factor is similar to the
nucleon axial form factor for $0.3 \ge x \ge 0.05$, and which
suggested that a large part of the observed increase of the $t$--slope
of the $J/\psi$ photoproduction cross section between fixed--target
energies ($E_{\gamma} \le 100\, {\rm GeV}, x \ge 0.05$) and HERA
energies ($x\le 10^{-2}$) is due to the contribution of the pion
cloud.
%
%

In the results shown in Figs.~\ref{fig_fpib2} and \ref{fig_conv}, the
distributions of pions in the nucleon $f_{\pi N}$ and $f_{\pi\Delta}$
were evaluated using the leading asymptotic expressions at large $b$,
Eqs.(\ref{f_pi_N_asymp}) and (\ref{f_pi_Delta_asymp}), which contain
only the leading power of $1/(\kappa b)$ in the pre-exponential
factor.  In principle, since we are integrating the distributions over
$b$ from a lower cutoff 
$b_0 = \langle b^2 \rangle_{\rm bulk}^{1/2} = 0.55\, {\rm fm}$, which
is numerically comparable to (in fact, even smaller than) 
$1/(2 M_\pi) = 0.73 \, {\rm fm}$, subleading terms of order 
$1/(\kappa b)$ in the pre-exponential factors, which become $\sim 1/(2
M_\pi b)$ at small $y$, could become important.  To estimate the
influence of these terms we have evaluated the integrals
(\ref{f_pi_b2_int}) also with the full expression for $f_{\pi N}$
derived from the pion exchange diagrams, Eq.(\ref{f_pi_N_full}), and
the corresponding result for $f_{\pi\Delta}$ obtained by numerical
evaluation of Eq.(\ref{f_pi_Delta_gauss}). We find the $b^2$--weighted
integral (\ref{f_pi_b2_int}) in this case to be $\sim 30\%$ larger for
the $N$, and $\sim 10\%$ larger for the $\Delta$ contribution,
resulting in a $\sim 20\%$ larger total value of 
$\langle b^2 \rangle_{\rm cloud}$ at $x \ll M_\pi / M_N$ than the one
shown in Fig.~\ref{fig_conv} (solid line).  The small difference
compared to the results obtained with the leading asymptotic
expressions for the distributions at large $b$ clearly demonstrates
that the effect we are considering is indeed due to large transverse
distances, where our approximations are justified.  Note that this is
because we are considering integrals over the distribution of pions
{\it weighted with} $b^2$, which strongly enhances contributions from
large distances; it would not be the case for the simple integral of
the distribution of pions over $b$ (the number of pions with 
$b > b_0$) which would enter in the pion cloud contribution to the
parton density, such as the flavor asymmetry of the sea quark
distributions, {\it cf.}\ the discussion in Section~\ref{sec_largenc}.
To conclude, we emphasize that the uncertainty discussed in this
paragraph concerns only the height of the ``jump'' of 
$\langle b^2 \rangle_{\rm cloud}$ between $x > M_\pi / M_N$ and 
$x \ll M_\pi / M_N$; the basic mechanism of suppression of the pion
cloud at $x > M_\pi / M_N$ is due to the exponential factor in the
asymptotic expansion at large $b$, and thus more robust.
%
%

In the large--$N_c$ limit, the sum of the $N$ and $\Delta$
contributions to $\langle b^2 \rangle_{\rm cloud}$ would be exactly 3
times larger than the $N$ contribution alone. As one sees from
Fig.~\ref{fig_conv}, our result for the sum of $N$ and $\Delta$
contributions with {\it physical} $N$ and $\Delta$ masses (solid line)
lies below 3 times $N$ only (dashed line) by a factor of $\sim 2/3$.
It thus appears as if the large--$N_c$ limit overestimated the
``exact'' result. However, there is the question which value one
should take for the common $N$ and $\Delta$ mass in the large--$N_c$
limit.  Multiplying the $N$ contribution in Fig.~\ref{fig_conv} by 3
amounts to taking $M_\Delta = M_N = 940\, {\rm Mev}$. This is an
extreme choice, and the resulting $\langle b^2 \rangle_{\rm cloud}$
should clearly be seen as an upper limit. Choosing larger values for
the the common mass in the large--$N_c$ limit, as are suggested by
most chiral soliton models, would lower the large--$N_c$ estimate
considerably, bringing it more in line with our ``exact'' result.

In the derivation of the large--$b$ asymptotics of the gluon density
in the nucleon, Eq.(\ref{convolution}), we neglected the ``intrinsic''
transverse size of the gluon distribution in the pion. Of course, the
transverse size of the distribution in the pion is expected to grow at
small $z = x/y$ (momentum fraction of the gluon relative to the pion).
This does not play any role in the region of the onset of the growth
of the transverse size of the nucleon, $x \sim M_\pi / M_N$, since the
configurations responsible for the growth of the nucleon size are
those at the lower limit of the $y$--integral, $y \sim x$ for which 
$z \sim 1$. However, it becomes an issue for values of $x$
considerably smaller than $M_\pi / M_N$. In other words, the finite
size of the pion does not change the onset of the growth of the
nucleon size in $x$, but may increase the height of the jump.

One could think of applying the reasoning of Section~\ref{sec_large}
to the pion itself and consider the growth of the transverse size of
the pion due to soft--pion exchange in the $t$--channel.  In the case
of the pion the growth of the transverse size should start already for
values $z \sim 1$, contrary to the nucleon, where it is ``postponed''
until $x \ll M_\pi / M_N$.  However, since the pion--pion coupling is
weaker than the pion--nucleon coupling the relevance of the ``pion
cloud of the pion'' for the transverse size of its gluon distribution
is questionable. In particular, in the large--$N_c$ limit the
pion--pion interaction is suppressed relative to the pion--nucleon
interaction. The transverse size of the pion is thus likely to be
determined by its quark core and can be studied only in models.  A
rough estimate of the transverse size of the pion is provided by its
electromagnetic charge radius [{\it cf.}\ Eq.(\ref{bulk_axial})]
\beq
\langle b^2 \rangle_\pi \;\; \approx \;\; \frac{2}{3} \,
\langle r^2 \rangle_\pi \;\; = \;\; 0.3 \, {\rm fm}^2 .
\eeq
This suggests that the size of the gluon distribution in the pion is
comparable to that of the bulk of the gluon distribution in the
nucleon. The finite size of the pion gives a positive contribution to
$\langle b^2 \rangle_{\rm cloud}$, which in principle should be
added. However, this would exceed the accuracy of our simple
two--scale picture in which the core radii (either nucleon or pion)
are parametrically smaller than $1/M_\pi$. A quantitative estimate of
the effects of the finite transverse size of the pion can be made
within the chiral quark--soliton model of the nucleon, which describes
the structure of the pion and the nucleon in a unified way
\cite{Diakonov:1996sr,Petrov:1998kf}. This approach would also
incorporate the effects of the pion--nucleon form factors in a
consistent way.
\section{Pion knockout in hard exclusive processes}
\label{sec_knockout}
We now turn to the question whether the pion cloud contributions to
the gluon density in the nucleon at large impact parameters could
directly be observed in exclusive photo/electroproduction experiments,
and whether in this way one could extract information about the gluon
distribution in the pion.

One possibility would be to measure the $t$--dependence of the
cross section for $\rho$ electroproduction off the nucleon, 
$\gamma^\ast + N \rightarrow \rho + N$, or for 
$J/\psi$-- or $\Upsilon$ photoproduction, over a sufficiently wide 
range such that one can restore the $b$--dependence for
$b \gtrsim 1/(2 M_\pi) \approx 0.7 \, {\rm fm}$.
(A similar program was carried out in Ref.\cite{Munier:2001nr} 
in the framework of the dipole picture of high--energy scattering.)
In this way one could extract the gluon density in the nucleon at
large $b$, which could be compared with the asymptotic expression
Eqs.(\ref{convolution}), (\ref{f_pi_N_asymp}) and (\ref{kappa_N}). 
A problem with this approach is that the theoretical prediction
Eq.(\ref{convolution}) has the form of a convolution in the momentum 
fraction of the pion, $y$, which would render the extraction of the 
gluon distribution in the pion difficult. Another problem is the
unknown size of sub-asymptotic contributions to the nucleon's
gluon distribution at large $b$, which have a faster exponential 
fall-off than the $N$ contribution but may be enhanced by numerical
prefactors. On the practical end, the resolution in $t$ at small $t$, 
and the $t$--range covered by the present HERA experiments, are definitely 
insufficient for making the transformation to $b$--space with the 
required accuracy. Hopefully, the necessary resolution could be reached 
at the planned Electron--Ion Collider (EIC) \cite{EIC}.

\begin{figure}[t]
\begin{center}
\begin{tabular}{c}
\psfrag{Nxp}{{\Large $N(p)$}}
\psfrag{Nyp}{{\Large $N(p')$}}
\psfrag{Pxk}{{\Large $\pi (k')$}}
\psfrag{Axl}{{\Large $a (l')$}}
\psfrag{k1}{{\Large $k$}}
\psfrag{q1}{{\Large $q$}}
\psfrag{t1}{{\Large $t$}}
\includegraphics[width=7.0cm,height=5.9cm]{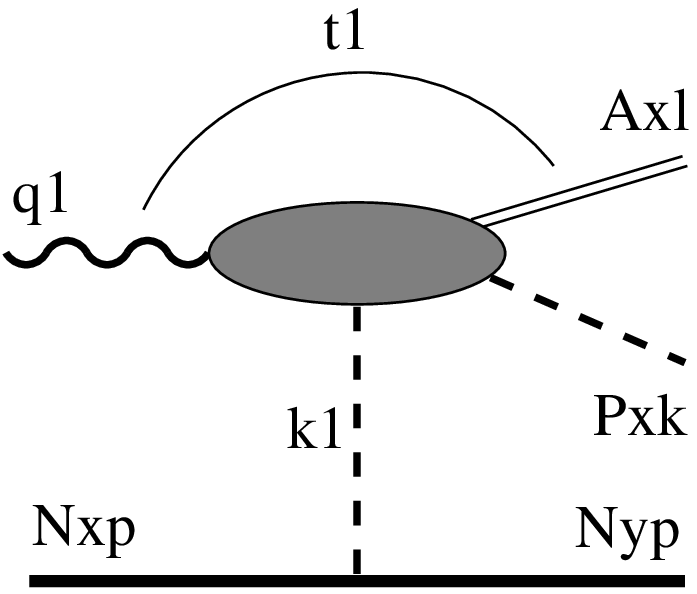}
\\[1ex] (a) \\[8ex]
\includegraphics[width=3.5cm,height=1.7cm]{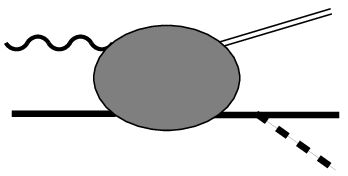}
\hspace{2em}
\includegraphics[width=3.5cm,height=1.7cm]{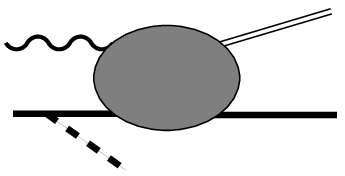}
\\[1ex] (b) 
\end{tabular}
\end{center}
\caption[]{(a) The mechanism for hard exclusive photo-- or 
electroproduction of a system $a$ (vector meson, heavy quarkonium), 
accompanied by wide--angle pion production, 
$\gamma + N \; \rightarrow \; a + \pi + N' \; (\mbox{or} \; \Delta)$. 
The blob denotes the amplitude for the subprocess 
$\gamma + \pi \rightarrow  a + \pi$.
(b) Competing reaction mechanisms in which the whole nucleon participates
in the hard process, and the pion is emitted from the external baryon 
lines. These contributions can be suppressed if the pion is emitted
under sufficiently large transverse momentum.}
\label{fig_knockout}
\end{figure}

A more direct measurement of the glue in the pion cloud is
possible in exclusive reactions in which the ``struck'' pion is observed 
in the final state. Here we wish to consider processes of the type 
\beq
\gamma^* \; + \; N \;\; \rightarrow \;\; a \; +\; \pi
\; + \; N' \; (\mbox{or} \; \Delta) ,
\label{knockout}
\eeq
in the kinematical region where the subprocess 
$\gamma^* + \pi \rightarrow a + \pi$ satisfies factorization theorems
for exclusive processes, see Fig.~\ref{fig_knockout}a. This could be
either a large--$Q^2$ process, with the produced particle $a$ a real
photon or a vector meson (in the latter case only the amplitude for
longitudinal polarization of the virtual photon should be considered),
or a process in which a heavy quarkonium is produced 
($a = J/\psi, \Upsilon$).  The process (\ref{knockout}) is an example
of a general class of reactions in which only a ``white cluster'' of
configurations in the target nucleon actually participate in the hard
scattering. It was pointed out in Ref.~\cite{Frankfurt:2002kz} that
such processes can be used to obtain qualitatively new information
about baryon structure.

To describe the process (\ref{knockout}) we choose $p$ and $q$
collinear and along the 3--axis. The outgoing particles are then
characterized by their longitudinal momentum fractions and transverse
momenta.  The momentum of the exchanged pion is completely fixed by
the external momenta and thus measurable; in particular, 
$k^2 = (p' - p)^2$. We characterize the momentum of the exchanged pion
by its longitudinal momentum fraction $y$ and transverse momentum
${\bf k}_{\perp}$. The amplitude for the process (\ref{knockout}) is
given by the product of the the amplitude to find the pion in the
nucleon with momentum fraction $y$ and transverse momentum 
${\bf k_\perp} = - {\bf p}'_\perp$, and the amplitude of the process
$\gamma + \pi \rightarrow a + \pi$.  At sufficiently small Bjorken $x$
the latter process is dominated by the gluon distribution in the
pion. The $t$--dependence of the latter is given by the two--gluon
form factor of the pion, $\Gamma_\pi (t)$.
%
%
As we argued already in Section~\ref{sec_large}, the characteristic
scale of this form factor, $m_{2g}^2$, is comparable to that of the
pion electromagnetic form factor, and the analogy with the nucleon
\cite{Frankfurt:2002ka} suggests that it may be as large as 
$m_{2g}^2 \approx 1 \, {\rm GeV}^2$, see Eq.(\ref{twogluon_pi}) and
after. In any event, the characteristic scale of this form factor is
{\it much harder} than the transverse momenta which could be generated
in the pion--baryon cluster system. This is crucial for our
discussion, since it guarantees that the reaction mechanism of
interest, Fig.~\ref{fig_knockout}a, can be separated from the
competing mechanism in which the whole nucleon is involved in the hard
process, and the extra pion is emitted by the incoming or outgoing
nucleon, see Fig.~\ref{fig_knockout}b. Keeping $(p - p')^2$ small, of
order $M_\pi^2$, and choosing $t = (l' - q)^2$ sufficiently large
($|{\bf l}_\perp| \geq 0.3 \ldots 0.4\, {\rm GeV}$), one effectively
suppresses contributions of the type of Fig.~\ref{fig_knockout}b.

We can write the the modulus squared of the amplitude of the process 
(\ref{knockout}) in the form
\be
|M_{\gamma + N \rightarrow a + \pi + N'}|^2 &=& \frac{4\pi}{y} 
\; w_{N \rightarrow N'} \; d_{\pi N} (y, {\bf k}_\perp)
\nonumber \\
&\times & |M_{\gamma + \pi \rightarrow a + \pi}|^2 .
\ee
Here $w_{N \rightarrow N'}$ is an isospin factor; its value is 
1 for $p \rightarrow p$ or $n \rightarrow n$, and 2 for
$p \rightarrow n$ or $n \rightarrow p$ transitions. Furthermore,
$M_{\gamma + \pi \rightarrow a + \pi}$ denotes the invariant
amplitude of the subprocess $\gamma + \pi \rightarrow a + \pi$. The
function $d_{\pi N} (y, {\bf k}_\perp)$ is defined as, up to a factor, 
the modulus squared of the amplitude for the nucleon to emit a pion, 
including the pion propagator, averaged (summed) over the incoming
(outgoing) nucleon helicities:
\beq
\frac{4\pi}{y} \; d_{\pi N} (y, {\bf k}_\perp)
\;\; \equiv \;\;
\frac{1}{2} \sum_{\lambda\lambda'} 
\frac{g_{\pi NN}^2 \; |\bar u' i \gamma_5 \tau^a u |^2 }
{(k^2 - M_\pi^2 )^2}.
\eeq
Explicit calculation gives
\beq
d_{\pi N} (y, {\bf k}_\perp)
\;\; = \;\; g_{\pi NN}^2 
\frac{s}{(s + M_\pi^2 )^2}
\label{d_pi_N_res} ,
\eeq
where $s \equiv (y^2 M_N^2 + {\bf k}_\perp^2)/\bar y = -k^2$ is the
spacelike pion virtuality. This function can be interpreted as the
transverse momentum--dependent distribution of pions in the nucleon.
Formally, the integral of Eq.(\ref{d_pi_N_res}) over all 
${\bf k}_\perp$, multiplied by 3 to account for the isospin
degeneracy, gives the total distribution of pions in the nucleon,
$H_\pi (y, -{\bf\Delta}_\perp^2 = 0)$, {\it cf.}\
Eq.(\ref{H_pi_N_kperp}) for ${\bf\Delta}_\perp = 0$.  Again, it is
physically more sensible to consider the distribution of pions
(\ref{d_pi_N_res}) in coordinate space. We define the Fourier
transform of Eq.(\ref{d_pi_N_res}) as
\be
\tilde d_{\pi N} (y, {\bf a}_\perp)
& \equiv &
\int\frac{d^2 k_\perp}{(2\pi)^2} \; 
e^{i ({\bf k}_\perp {\bf a}_\perp)}\;
d_{\pi N} (y, {\bf k}_\perp) ,
\ee
where ${\bf a}_\perp$ is a transverse coordinate variable conjugate to
${\bf k}_\perp$. At large $|{\bf a}_\perp |$ it behaves as
\be
\tilde d_{\pi N} (y, {\bf a}_\perp)
&\sim & 
\frac{g_{\pi NN}^2 \, M_\pi^2}{16 \sqrt{2\pi^3}} \; \bar y \;
\sqrt{\frac{|{\bf a}_\perp|}{\tilde\kappa_N}} \; 
e^{\displaystyle -\tilde\kappa_N |{\bf a}_\perp|}, 
\ee
where
\beq
\tilde\kappa_N \;\; \equiv \;\; \sqrt{y^2 M_N^2 + \bar y M_\pi^2} .
\eeq
For $y \sim M_\pi / M_N$ the width of this distribution in transverse
space is of the order $1/M_\pi$, {\it i.e.}, much larger than the size
of typical configurations in the nucleon, $1/M_N$.  In this region the
exponential dependence could in principle be observed experimentally
in the Fourier transform of the cross section of the reaction
(\ref{knockout}) with respect to ${\bf k_\perp} = - {\bf p'}_\perp$,
{\it i.e.}, the transverse momentum of the outgoing nucleon.  Due to
the broad distribution of the cross section in $t$ it should be easy
to make a cut on $-t_{\rm min}$ which would allow to measure 
$d_{\pi N} (y,{\bf k_\perp} )$ in a sufficiently wide range of 
${\bf p'}_\perp$ such that one could reconstruct 
$\tilde d_{\pi N} (y,{\bf a}_\perp)$.  Alternatively, one could model
$\tilde d_{\pi N} (y,{\bf a}_\perp)$ in the whole range of 
$|{\bf a}_\perp|$, calculate the Fourier transform, and check how
small ${\bf p}'_\perp$ needs to be in order for the region of 
$|{\bf a}_\perp|$ which we consider as safe to give the dominant
contribution.  We assume here that for 
$|{\bf k}_\perp| \ll |{\bf k}_\perp'|$ the dependence of the amplitude
of the subprocess $\gamma + \pi \rightarrow a + \pi$ on the transverse
momentum of the ``initial'' pion, ${\bf k}_\perp$, can be neglected.

At sufficiently small Bjorken $x$ the hard subprocess 
$\gamma + \pi \rightarrow a + \pi$ is dominated by the gluon
distribution in the pion. Measuring the cross section of the process
$\gamma + N \rightarrow a + \pi + N'$, and isolating the contribution
from pion exchange as described above, one could thus measure the
gluon distribution in the pion, including its $t$--dependence
(two--gluon form factor).  The complete discussion of the cross
section of this process, appropriate kinematical cuts, {\it etc.}\ is
beyond the scope of the present investigation and will given
elsewhere.  Here we note only that at sufficiently small $x$ a simple
expression can be written for the ratio of the squared amplitudes of
the process (\ref{knockout}) and the corresponding production process
without pion emission:
\be
\frac{|M_{\gamma + N \rightarrow a + \pi + N'}|^2}
{|M_{\gamma + N \rightarrow a + N}|^2}
&=&
\frac{4\pi}{y}
\; w_{N \rightarrow N'} \; d_{\pi N} (y, {\bf k}_\perp)
\nonumber \\
&\times & \left[ \frac{\displaystyle (x/y) \; g_\pi (x/y, Q^2)}
{\displaystyle x g(x, Q^2)}\right]^2
\nonumber \\
&\times &
\left[ \frac{\displaystyle \Gamma_\pi (t)}
{\displaystyle \Gamma (t)} \right]^2 .
\ee
Here $Q^2$ denotes the scale appropriate for the hard process, see
Refs.\cite{Frankfurt:1997fj,Frankfurt:1995jw} for a discussion, and
$\Gamma_\pi (t)$ and $\Gamma (t)$ are the two--gluon form factors of
the pion and nucleon, respectively (in both processes $t$ denotes the
invariant momentum transfer between the incoming photon and the
outgoing vector meson/photon, see Fig.~\ref{fig_knockout}a).  Since
the distribution of pions is not too small for small $y$, and the
gluon density in the pion strongly increases with decreasing $x/y$, we
expect the cross section for the process (\ref{knockout}) to be
non-negligible for $y\sim 0.05\ldots 0.1$. In addition, if indeed the
two gluon form factor of the pion decreased slowly with $t$ as in
Eq.~(\ref{twogluon_pi}), the production rate would decrease with
$|{\bf k}'_\perp |$ only as $\propto 1/|{\bf k'_\perp}|^2$.  One may
thus expect noticeable rates for pions produced with transverse
momenta $|{\bf k}'_\perp | \geq 2 \, {\rm GeV}$, which is within the
acceptance of the current HERA detectors.
\section{Conclusions and outlook}
\label{sec_conclusions}
In this paper we have studied the role of the pion cloud of the
nucleon for the gluon density at large transverse distances.  We have
identified a specific contribution to the gluon density of transverse
size $1/M_\pi$, due to soft--pion exchange, which becomes ``visible''
for $x \ll M_\pi / M_N$. This results in a finite increase of the
average transverse size of the nucleon as $x$ drops significantly
below $M_\pi / M_N$, in agreement with the observed change of the
$t$--dependence of the $J/\psi$ photoproduction cross section with
incident energy.  While this phenomenon has been described previously
at the qualitative level, the present investigation shows that with
the phenomenological parameters for soft--pion exchange one indeed
obtains a consistent quantitative picture.

Let us summarize again in which range of $x$ the soft--pion exchange
contribution to the gluon density is meaningful. We have seen that
this mechanism sets in for $x < M_\pi / M_N \sim 10^{-1}$, when the
relativistic kinematics allows the pions to propagate distances of
order $1/M_\pi$ in the transverse plane. It grows to its full strength
for $x \lesssim 10^{-2}$, when the momentum fraction of the gluons
relative to the pions becomes small ($\sim 10^{-1}$). In principle
this contribution to the gluon density survives down to considerably
smaller values of $x$.  However, for such values of $x$ other effects
such as multi-step Gribov diffusion become dominant.

The gluon density at large transverse distances is an important
element of the structure of the nucleon which can be calculated
theoretically in a model--independent fashion. It can be probed in
exclusive processes such as $J/\psi$ photoproduction, or hard
electroproduction of real photons or vector mesons at small $x$, by
isolating contributions from large impact parameters through
measurement of the $t$--dependence. The QCD factorization theorem
guarantees that this feature of the nucleon is process--independent as
long as the conditions for factorization are met. A particularly
promising way of accessing the glue at large transverse distances
appears to be hard exclusive production off the nucleon with
associated ``pion knockout'', which should be possible to observe at
HERA.

Our results for the asymptotic behavior of the gluon density at large
transverse distances provide model--independent constraints for
parametrizations or dynamical models of the impact parameter
dependence of the nucleon's gluon density. Such parametrizations are
needed to describe {\it e.g.}\ diffractive Higgs production in
proton--proton collisions at LHC energies \cite{FELIX}.

The techniques developed in the present paper can readily be extended
to quark distributions, including polarized distributions. The
large--$b$ limit of these distributions is in principle also due to
soft--pion exchange, and can be described in a model--independent
way. However, the minimum number of exchanged pions in the
$t$--channel depends on the $G$--parity of the operator measuring the
parton distribution, and the relevance of the asymptotic region for
eventual observable physical phenomena needs to be studied
case--by--case. The impact parameter representation of parton
distributions is also useful in connection with specific dynamical
models of the nucleon, as it leads to new insights into the dynamical
origin of these quantities. For example, with regard to the flavor
asymmetry in the nucleon's sea quark distributions, 
$\bar u (x) - \bar d (x)$ and $\Delta\bar u (x) - \Delta\bar d (x)$,
it may help to resolve the long-standing issue to which extent these
asymmetries are due to the pion cloud of the nucleon, or to Pauli
blocking by the valence quarks \cite{Fries:2002um}.

In the present investigation we have limited ourselves to the
generalized parton distributions at zero longitudinal momentum
transfer (zero ``skewedness''), which is sufficient for studying the
impact parameter dependence of the usual parton densities in the
nucleon.  The inclusion of skewedness effects is an important problem
which we leave for future treatment. In particular, this would allow
to include into the considerations a much wider range of experiments,
such as deeply--virtual Compton scattering or meson production
(pseudoscalar and vector) in the valence region. First steps have been
taken to extend the impact parameter representation to generalized
parton distributions at non-zero skewedness \cite{Diehl:2002he};
however, one is still lacking a simple interpretation as is available
in the diagonal case.
\par
We are grateful to H.~Weigert for inspiration and help during the initial 
stages of this work, and to M.~Burkardt, L.~L.~Frankfurt, A.~Freund, N.~Kivel, 
G.~A.~Miller, M.~V.~Polyakov, A.~V.~Radyushkin, A.~Sch\"afer, W.~S\"oldner, 
and M.~Stratmann for interesting discussions. We thank the Institute for
Nuclear Theory at the University of Washington for its hospitality
during the time in which this work was completed. M.~S.\ is an 
Alexander--von--Humboldt Fellow. C.~W.\ is a Heisenberg Fellow 
(Deutsche Forschungsgemeinschaft). This work has been supported by D.O.E.
\newpage
\appendix
\section{Distribution of pions in the nucleon at large impact parameters}
\label{app_twopion}
\subsection{Evaluation of the pion exchange graphs}
\label{app_feynman}
In this appendix we outline the derivation of the asymptotic behavior
of the ``parton distribution'' of pions in the nucleon at large impact
parameters. We first consider the distributions obtained with $N$
intermediate states only; the $\Delta$ contributions will be included
subsequently.

In the limit of large transverse distances it is justified to treat
the nucleon and the pion as pointlike elementary particles. We assume
a pion--nucleon coupling of pseudoscalar form
\be
L_{\pi NN} \;\; = \;\; 
g_{\pi NN} \bar N i \gamma_5 \tau^a \pi^a N ,
\ee 
where $\tau^a$ are the isospin Pauli matrices.  The result for the
large--$b$ behavior is in fact independent of the type of coupling;
{\it i.e.}\ it is the same for all couplings which are equivalent in
the soft--pion limit.

We first evaluate the $t$--dependent distributions of pions in the
nucleon, as defined by Eq.(\ref{H_pi_E_pi_def}). The nucleon matrix
element of the pion light--ray operator is computed using standard
Feynman rules,
\be
&& \langle N(P + \Delta/2) |\; \pi^a (-\lambda n/2) \pi^a (\lambda n /2)
\; | N(P - \Delta /2) \rangle
\nonumber \\
&=& 3 i g_{\pi NN}^2 \; 
\int\frac{d^4 k}{(2\pi )^4} \; \left[ 
e^{-i \lambda (kn)} + e^{i \lambda (kn)} \right]
\nonumber \\
&\times& 
\frac{1}{
\left[ (k + \Delta/2)^2 - M_\pi^2 + i0 \right]
\left[ (k - \Delta/2)^2 - M_\pi^2 + i0 \right]}
\nonumber \\
&\times& 
\frac{\bar u' \; i \gamma_5 \; (\hat P - \hat k + M_N) 
\; i \gamma_5 \; u}
{\left[ (P - k)^2 - M_N^2 + i0 \right]} .
\label{me_feynman_mom}
\ee
The factor 3 arises from the summation over the pion isospin
projection.  In order to read off the expressions for the distribution
functions we have to convert the Dirac structures in the numerator to
the form of the L.H.S.\ of Eq.(\ref{H_pi_E_pi_def}). The $\gamma_5$
matrices can be eliminated using the anticommutation relations. The
integral over the $\hat k$ term can be projected on the structures
$\hat P, \hat \Delta$ and $\hat n$ by appropriate projections of the
four--vector $k$. Finally, we use the identities 
$\bar u' \hat P u = M_N \bar u' u$ and $\bar u' \hat \Delta u = 0$, 
as well as
\beq
\frac{P_\nu}{M_N} \bar u' u
\;\; = \;\; \bar u' \gamma_\nu u
- \frac{1}{2 M_N} \bar u' \sigma_{\mu\nu} \Delta^\mu u ,
\eeq
which follow from the Dirac equation for the external nucleon spinors.
The net result is that we can replace in Eq.(\ref{me_feynman_mom})
\be
&& \bar u' \; i \gamma_5 \; (\hat P - \hat k + M_N) 
\; i \gamma_5 \; u
\nonumber \\
&\rightarrow &
\frac{1}{(Pn)} \left[ -(kP) 
\; + \; (P^2 - M_N^2) \frac{(kn)}{(Pn)} \right] \bar u' \hat n u
\nonumber \\
&+& \frac{M_N (kn)}{2 (Pn)^2} \bar u' \sigma_{\mu\nu} 
\Delta^\mu n^\nu u .
\ee
To get the distribution of pions as a function of the momentum
fraction $y$ we still have to compute the Fourier transform of
Eq.(\ref{me_feynman_mom}) with respect to the longitudinal distance
variable $\lambda$, see Eq.(\ref{H_pi_E_pi_def}). The integral of the
combined exponential factors results in delta functions,
\beq
\int_{-\infty}^\infty
\frac{d\lambda}{2\pi} \; e^{i \lambda y (Pn)} \; e^{\mp i \lambda (kn)}
\;\; = \;\; \delta\!\left[ y (Pn) \mp (kn) \right] .
\eeq
Combining everything, we see that the distribution of pions resulting
from $N$ intermediate states is given in covariant form by
\be
H_{\pi N} (y, t)
&=& 6 \, i g_{\pi NN}^2 \; (Pn) y \; 
\int\frac{d^4 k}{(2\pi )^4} \delta\!\left[ y (Pn) - (kn) \right]
\nonumber \\
&\times& 
\frac{1}
{\left[ (k + \Delta/2)^2 - M_\pi^2 + i0 \right]}
\nonumber \\
&\times&
\frac{1}
{\left[ (k - \Delta/2)^2 - M_\pi^2 + i0 \right]}
\nonumber \\
&\times& 
\frac{1}
{\left[ (P - k)^2 - M_N^2 + i0 \right]}
\nonumber \\
&\times& \left[ -(kP) \; 
\; - \; \frac{(kn)}{(Pn)} \, \frac{\Delta^2}{4} \right] 
\nonumber \\
&-& (y \rightarrow -y) ,
\label{H_pi_N_covariant}
\ee
where $t = \Delta^2$. For the helicity--flip distribution 
$E_{\pi N} (y, t)$ the expression in brackets in the next--to--last
last line should be replaced by $M_N^2 (kn)/(Pn)$. In the following we
quote only the results for the distribution $H_{\pi N}$ used in the
present investigation; the corresponding expressions for $E_{\pi N}$
can be derived analogously.

To evaluate the momentum integral we choose the frame described in 
Section~\ref{sec_npd}, where $n^\mu = (1, 0, 0, -1)$, 
$nP = P^+$, and ${\bf P}_\perp = 0$, see Eqs.(\ref{P_plus_def}),
which implies $P^- = P^2/P^+$. Furthermore, $\Delta^+ = \Delta^- = 0$,
and thus ${\bf\Delta}_\perp^2 = -t$. Corresponding
light-like components are introduced for the loop momentum, $k$.
The integral over $k^+$ is taken trivially; the delta function restricts
$k^+$ to $y P^+$. The integral over $k^-$ is then performed using 
Cauchy's theorem. The poles of the three propagators are located at
\be
k^- &=& \frac{({\bf k}_\perp \pm {\bf \Delta_\perp}/2)^2 + M_\pi^2 - i0}
{y P^+} , 
\\
k^- &=& \frac{-{\bf k}_\perp^2 - y M_N^2 
+ \bar y {\bf\Delta}_\perp^2 / 4 + i0}
{\bar y P^+} .
\label{pole_nucleon}
\ee
The integral is non-zero only if the poles lie on different sides of 
the real axis, which implies $y \bar y > 0$, or $0 < y < 1$.
Closing the contour in the upper half-plane, encircling the pole
(\ref{pole_nucleon}), we obtain for $0 < y < 1$
\be
H_{\pi N} (y, -{\bf\Delta}_\perp^2 )
&=& \frac{3 g_{\pi NN}^2}{8\pi} \; \frac{y}{\bar y} \; 
\int\frac{d^2 k_\perp}{(2\pi )^2} 
\nonumber \\
&\times& 
\frac{s_+ + s_- - \bar y {\bf\Delta}_\perp^2}
{(s_+ + M_\pi^2) (s_- + M_\pi^2)} ,
\label{H_pi_N_kperp}
\ee
where
\be
s_{\pm} &\equiv& \frac{1}{\bar y} \left[ 
({\bf k}_\perp \pm \bar y {\bf\Delta}_\perp / 2)^2
+ y^2 M_N^2 \right]
\nonumber \\
&=& -(k \pm \Delta /2)^2 \hspace{2em}
\mbox{at the pole} 
\label{s_pm}
\ee
are the pion virtualities. The minimum value of the pion virtuality in 
the loop is
\beq
s_{{\rm min}, N} \;\; = \;\; \frac{y^2 M_N^2}{\bar y} .
\label{s_min_N_app}
\eeq
For ${\bf\Delta}_\perp = 0$ our result (\ref{H_pi_N_kperp}) reduces to
the well--known expression for the isoscalar distribution of pions in
the meson cloud model, $f_{\pi N}(y)$ of Ref.\cite{Koepf:1995yh}, up
to form factors at the pion--nucleon vertices, which are not needed in
our approach, see below. Note that in Ref.\cite{Koepf:1995yh} the
isospin factor $3$ is included in the definition of $g_{\pi NN}$.

In Eq.(\ref{H_pi_N_kperp}) it is assumed that $0 < y < 1$; the
distribution for $-1 < y < 0$ follows trivially from
$H_{\pi N} (-y, -{\bf\Delta}_\perp^2 ) = 
-H_{\pi N} (y, -{\bf\Delta}_\perp^2 )$.
In the following, for simplicity, we shall always assume that 
$0 < y < 1$, and that the distribution at negative $y$ is
restored in this way.

The transverse momentum integral in Eq.(\ref{H_pi_N_kperp}) for the
distribution $H_\pi (y, -{\bf\Delta}_\perp^2)$ is logarithmically
divergent. However, this divergence is irrelevant for the large--$b$
behavior of the corresponding impact parameter--dependent
distribution. It could be removed by subtracting the function at
$-{\bf\Delta}_\perp^2 = t = 0$,
\beq
H_{\pi N} (y, -{\bf\Delta}_\perp^2 ) \;\; \rightarrow \;\; 
H_{\pi N} (y, -{\bf\Delta}_\perp^2 ) - H_{\pi N} (y, 0) .
\eeq
In impact parameter space this would amount to subtracting a delta
function at ${\bf b} = 0$, which is irrelevant at large $b$.  With the
saddle point method employed below, we do not need to perform this
subtraction explicitly.

For deriving the large--$b$ behavior of the corresponding impact
parameter--dependent distributions it is convenient to introduce the
integral representation
\beq
\frac{1}{A} \;\; = \;\; \int_0^\infty \! d\alpha\; e^{-\alpha A}
\hspace{3em} (A > 0)
\label{exponential_integral}
\eeq
for the ${\bf k}_\perp$--dependent denominators in
Eq.(\ref{H_pi_N_kperp}).  The integral over ${\bf k}_\perp$ then
reduces to a Gaussian integral, which can be computed. We obtain
\be
\lefteqn{H_{\pi N} (y, -{\bf\Delta}_\perp^2 ) } && 
\nonumber \\
&=& \frac{3 g_{\pi NN}^2}{16\pi^2} \; y
\; \int_0^\infty \! d\alpha\; \int_0^\infty \! d\beta\;
\nonumber \\
&\times& \exp\left[ - (\alpha + \beta) (s_{{\rm min}, N} + M_\pi^2)
- \frac{\alpha\beta}{\alpha + \beta} \bar y {\bf \Delta}_\perp^2 \right]
\nonumber \\
&\times& \frac{1}{\alpha + \beta}
\left[ \frac{1}{\alpha + \beta} + s_{{\rm min}, N}
- \frac{\alpha\beta}{(\alpha + \beta )^2} \bar y 
{\bf\Delta}_\perp^2 \right] , 
\label{H_pi_N_gauss}
\ee
where $s_{{\rm min}, N}$ is defined in Eq.(\ref{s_min_N_app}).  We can
now pass to the impact parameter--dependent distribution of pions in
the nucleon, {\it cf.}\ Eq.(\ref{f_pi_N_def}). The Fourier transform
of Eq.(\ref{H_pi_N_gauss}) with respect to ${\bf\Delta}_\perp$ is
again a Gaussian integral. We get
\be
\lefteqn{f_{\pi N} (y, b)} && \nonumber \\
&=& \frac{3 g_{\pi NN}^2}{64\pi^3} \; \frac{y}{\bar y} 
\; \int_0^\infty \! d\alpha\; \int_0^\infty \! d\beta\;
\nonumber \\
&\times& \exp \left[ - (\alpha + \beta) (s_{{\rm min}, N} + M_\pi^2)
- \frac{\alpha + \beta}{\alpha\beta} \; \frac{b^2}{4\bar y} \right]
\nonumber \\
&\times& \frac{1}{\alpha\beta}
\left( s_{{\rm min}, N} 
+ \frac{1}{\alpha\beta} \; \frac{b^2}{4\bar y} \right) .
\label{f_pi_N_gauss}
\ee
In this representation the large--$b$ asymptotics can be studied using
the saddle point method. The stationary point of the exponent is at
\beq
\alpha \;\; = \;\; \beta \;\; = \;\; 
\frac{b}{2} \left[ \bar y \, (s_{{\rm min}, N} + M_\pi^2) \right]^{-1/2} ,
\eeq
and one immediately sees that with exponential accuracy the large--$b$ 
asymptotics is given by
\beq
f_{\pi N} (y, b) \;\; \sim \;\; e^{\displaystyle - \kappa_N b},
\eeq
with
\be 
\kappa_N &\equiv& 2 \left( \frac{s_{{\rm min}, N} 
+ M_\pi^2}{\bar y} \right)^{1/2} .
\label{kappa_N_from_s_min}
\ee
One can easily determine the pre-exponential factor by computing the
Gaussian integral over deviations from the stationary point; the
result is given in Eq.(\ref{f_pi_N_asymp}). Alternatively, one can
express the integral Eq.(\ref{f_pi_N_gauss}) in terms of modified
Bessel functions,
\be
f_{\pi N} (y, b) &=&
\frac{3 g_{\pi NN}^2}{16\pi^3} \; \frac{y}{\bar y} 
\left\{ s_{{\rm min}, N} \; \left[ K_0 (\kappa_N b/2) \right]^2 \right.
\nonumber \\
&+& \left. (s_{{\rm min}, N} + M_\pi^2 ) \; 
\left[ K_1 (\kappa_N b/2) \right]^2 
\right\} ,
\label{f_pi_N_full}
\ee
and recover the asymptotic behavior Eq.(\ref{f_pi_N_asymp}) by
substituting the asymptotic expressions for the Bessel functions at
large arguments.  In particular, in this way one can get also the
subleading corrections to the pre-exponential factor involving higher
powers of $1/(\kappa_N b)$, which are more difficult to compute using
the saddle point method.
\subsection{Large--$b$ behavior from $t$--singularities}
\label{app_cutkosky}
The large--$b$ behavior of the distribution of pions in the nucleon
can also be derived in an alternative way, which directly exhibits the
connection of the large--$b$ asymptotics with the singularities of the
pion and nucleon propagators in the Feynman integral
(\ref{H_pi_N_covariant}). Performing in Eq.(\ref{f_pi_N_def}) the
integral over the angle between the transverse vectors
${\bf\Delta}_\perp$ and ${\bf b}$ we obtain 
(here $\Delta_\perp \equiv |{\bf\Delta}_\perp|$)
\beq
f_{\pi N} (y, b) \;\; = \;\; 
\int_0^\infty \frac{d\Delta_\perp}{2\pi} 
\; \Delta_\perp \; J_0 (b\Delta_\perp ) \; 
H_{\pi N} (y, -\Delta_\perp^2 ) ,
\eeq
where $J_0$ denotes the Bessel function. Substituting the latter by
its asympto\-tic form for large arguments, 
$J_0 (b\Delta_\perp ) \sim (2\pi b \Delta_\perp )^{-1/2} 
\left[ e^{i (b\Delta_\perp - \pi /4)} +
e^{-i (b\Delta_\perp - \pi /4)}\right]$, 
and rotating the integration contour in the complex $\Delta_\perp$
plane to run along the positive (negative) imaginary axis in the
integral with the first (second) exponential factor, setting
$\Delta_\perp = \pm i \delta + 0$ with $\delta \geq 0$ real, we get
\be
f_{\pi N} (y, b) &=&  \frac{1}{\sqrt{2 \pi b}}
\; \int_{\mbox{\scriptsize threshold}}^\infty d\delta 
\; \sqrt{\delta} \; e^{-b \delta} \;
\nonumber \\
&\times& \frac{1}{\pi} \, \mbox{Im} \, H_{\pi N} (y, \delta^2 + i0) .
\label{large_b_from_im}
\ee
The large--$b$ behavior of the impact parameter--dependent
distribution is governed by the threshold behavior of the imaginary
part of the $t$--dependent distribution of pions, $H_{\pi N} (y, t)$,
which appears for some nonzero positive $t$. The imaginary part can
directly be computed from the Feynman integral
(\ref{H_pi_N_covariant}) using the Cutkosky rules, replacing the
propagators by appropriate delta functions. One finds that the
$t$--channel cut starts at $\delta = \kappa_N$, and that the leading
threshold behavior of the imaginary part for $\delta \rightarrow
\kappa_N + 0$ is
\be
\frac{1}{\pi} \, \mbox{Im} \, H_{\pi N} (y, \delta^2 + i0)
&=& \frac{3 g_{\pi NN}^2}{16 \pi^2} \;
\frac{y}{\bar y} \left( 2 s_{{\rm min}, N} + M_\pi^2 \right)
\nonumber \\
&\times& \left[ \frac{2}{\kappa_N^3 (\delta - \kappa_N )} \right]^{1/2} .
\ee
Substituting this in Eq.(\ref{large_b_from_im}) and computing the
integral over $\delta$, one obtains the asymptotic behavior quoted in
Eq.(\ref{f_pi_N_asymp}).
\subsection{The $\Delta$ contribution}
\label{app_Delta}
The calculation of the contribution from $\Delta$ intermediate states
proceeds along the same lines as that of the $N$ contribution; we only
outline the essential steps. We assume a $\pi N \Delta$ interaction
Lagrangian of the form
\be
L_{\pi N \Delta} &=& \frac{i g_{\pi N \Delta}}{\sqrt{2} M_N}
\left[ \bar p \; \partial_\mu \pi^- \; \Delta_\mu^{++} 
+ \sqrt{\frac{2}{3}} \; \bar p \; \partial^\mu \pi^0 \; \Delta^+_\mu
\right.
\nonumber \\
&+& \left. \frac{1}{\sqrt{3}} \; \bar p \; \partial^\mu \pi^+ \; \Delta^0_\mu
\;\; + \;\; \mbox{h.c.} \;\; + \ldots \right] ,
\label{L_pi_N_Delta}
\ee
where $p$ is the proton field (we skip the terms for the neutron), and
$\Delta_\mu$ the spin--3/2 Rarita--Schwinger field describing the
$\Delta$ resonance [not to be confused with the four--vector of the
momentum transfer in the matrix element (\ref{me_feynman_mom})].  Our
definition of the coupling constant $g_{\pi N \Delta}$ corresponds to
that of Ref.\cite{Adkins:1983ya}, and differs from the one used by
Koepf {\it et al.} \cite{Koepf:1995yh}, 
$(g_{\pi N \Delta})_{\mbox{\scriptsize Koepf et al.}}  = \sqrt{2} \;
(g_{\pi N \Delta})_{\mbox{\scriptsize this work}}$.
The relative coefficients of the terms in Eq.(\ref{L_pi_N_Delta}) are
dictated by isospin invariance. One sees that the total distribution
of pions in the proton ({\it i.e.}, the sum of distributions of
$\pi^+, \pi^-$ and $\pi^0$) resulting from $\Delta$ intermediate
states is equal to the $\pi^- \Delta^{++}$ contribution multiplied by
$(1 + 2/3 + 1/3) = 2$.

It the calculation of the Feynman graphs for the $\Delta$ contribution,
we use the propagator of the spin--3/2 Rarita--Schwinger field in the 
form ($l$ is the momentum)
\be
&& \frac{\hat l + M_\Delta}{l^2 - M_\Delta^2 + i0}
\left[ -g_{\mu\nu} + \frac{1}{3} \gamma_\mu \gamma_\nu
+ \frac{2}{3 M_\Delta^2} l_\mu l_\nu \right.
\nonumber \\
&& - \left. \frac{1}{3 M_\Delta} (l_\mu \gamma_\nu - \gamma_\mu l_\nu )
\right] .
\ee
After projecting on the Dirac structures appearing in the L.H.S.\ 
of Eq.(\ref{H_pi_E_pi_def}), in analogy to the $N$ case, we get
[{\it cf.}\ Eq.(\ref{H_pi_N_covariant})]
\be
\lefteqn{
H_{\pi\Delta} (y, t)} && 
\nonumber \\
&=& \frac{4 \, i g_{\pi N\Delta}^2}{3 \, M_N^2 \, M_\Delta^2} 
\; (Pn) y \; 
\int\frac{d^4 k}{(2\pi )^4} \delta\!\left[ y (Pn) - (kn) \right]
\nonumber \\
&\times& 
\frac{1}
{\left[ (k + \Delta/2)^2 - M_\pi^2 + i0 \right]}
\nonumber \\
&\times&
\frac{1}
{\left[ (k - \Delta/2)^2 - M_\pi^2 + i0 \right]}
\nonumber \\
&\times& 
\frac{1}
{\left[ (P - k)^2 - M_\Delta^2 + i0 \right]}
\nonumber \\
&\times& 
\left[ (P - k, k)^2 - (k\Delta )^2/4 + M_\Delta^2 (- k^2 + \Delta^2/4) 
\right]
\nonumber \\
&\times& 
\left[ -(kP) - \frac{(kn)}{(Pn)} \, \frac{\Delta^2}{4} 
+ M_N (M_N + M_\Delta ) \right]
\nonumber \\
&-& (y \rightarrow -y) .
\ee
For the helicity--flip distribution $E_{\pi\Delta} (y, t)$ the
expression in the bracket in the next--to--last line should be
replaced by
$\left[ M_N^2 (kn)/(Pn) - M_N (M_N + M_\Delta ) \right]$. 
Introducing light--like coordinates as above, and taking the integral
over $k^-$ using Cauchy's theorem, we finally get, for $0 < y < 1$
[{\it cf.}\ Eq.(\ref{H_pi_N_kperp})]
\be
\lefteqn{
H_{\pi\Delta} (y, -{\bf\Delta}_\perp^2 )} && 
\nonumber \\
&=& \frac{g_{\pi N\Delta}^2}{48 \pi \, M_N^2 \, M_\Delta^2} 
\frac{y}{\bar y} \int\frac{d^2 k_\perp}{(2\pi )^2} 
\frac{1}{(s_+ + M_\pi^2) (s_- + M_\pi^2)}
\nonumber \\
&\times& 
\left[ 2 M_\Delta^2 (s_+ + s_- - {\bf \Delta}_\perp^2 )
\right. 
\nonumber \\
&& \left. + \; (s_+ - M_\Delta^2 + M_N^2 ) (s_- - M_\Delta^2 + M_N^2 ) 
\right]
\nonumber \\
&\times&
\left[ 2 (M_\Delta + M_N)^2 + s_+ + s_- 
- \bar y {\bf\Delta}_\perp^2 \right] ,
\label{f_pi_Delta_kperp}
\ee
where now the pion virtualities are given by
\beq
s_{\pm} \;\; \equiv \;\; \frac{1}{\bar y} 
({\bf k}_\perp \pm \bar y {\bf\Delta}_\perp / 2)^2
+ s_{{\rm min}, \Delta} ,
\eeq
with
\beq
s_{{\rm min}, \Delta} \;\; \equiv \;\;
\frac{1}{\bar y} 
\left[ y^2 M_N^2 + y (M_\Delta^2 - M_N^2 ) \right] .
\eeq
For zero momentum transfer, ${\bf\Delta}_\perp = 0$,
Eq.(\ref{f_pi_Delta_kperp}) reduces to the $\Delta$ contribution to
the isoscalar distribution of pions in the meson cloud model quoted in
Ref.\cite{Koepf:1995yh} [{\it cf.}\ the comment on the different
conventions for $g_{\pi N\Delta}$ after Eq.(\ref{L_pi_N_Delta})], up
to form factors at the $\pi N \Delta$ vertices, which are not needed
in our approach. Again, we are interested in
Eq.(\ref{f_pi_Delta_kperp}) only as a means to derive the large--$b$
asymptotics of the corresponding impact parameter--dependent
distribution. This is done in complete analogy to the $N$ case, using
the integral representation (\ref{exponential_integral}) to convert
Eq.(\ref{f_pi_Delta_kperp}) to a Gaussian integral. For reference, we
quote the result for the $b$--dependent distribution, corresponding to
Eq.(\ref{f_pi_N_gauss}):
\be
\lefteqn{f_{\pi\Delta} (y, b)} && \nonumber \\
&=& \frac{g_{\pi N\Delta}^2}{384\pi^3 M_N^2 M_\Delta^2} \; \frac{y}{\bar y} 
\; \int_0^\infty \! d\alpha\; \int_0^\infty \! d\beta\;
\nonumber \\
&\times& \exp \left[ - (\alpha + \beta) (s_{{\rm min}, \Delta} + M_\pi^2)
- \frac{\alpha + \beta}{\alpha\beta} \; \frac{b^2}{4\bar y} \right]
\nonumber \\
&\times& \frac{1}{\alpha\beta}
\left\{ (X + A_+^2) \left[ (X + A_+^2) (X + A_-^2) \phantom{\frac{0}{0}}
\right. \right. 
\nonumber \\
&+& \left. \left. 
\frac{(\alpha + \beta )^2}{\alpha\beta} \, X \, A_+ A_- \right] 
+ \ldots \right\} ,
\label{f_pi_Delta_gauss}
\ee
where 
\be
X &\equiv& \frac{b^2}{4\alpha\beta\bar y} ,
\\
A_\pm^2 &\equiv& s_{{\rm min}, \Delta} + (M_\Delta \pm M_N)^2 
\nonumber \\
&=& \frac{(M_\Delta \pm \bar y M_N)^2}{\bar y} .
\ee
The ellipsis in Eq.(\ref{f_pi_Delta_gauss}) denotes terms which
produce only subleading corrections in $1/b$ to the pre-exponential
factor in the large--$b$ asymptotics, which we have not written out
for brevity.  The large--$b$ behavior can again be studied using the
saddle point method; the exponential decay constant is given by
Eq.(\ref{kappa_N_from_s_min}) with $s_{{\rm min, N}}$ replaced by
$s_{{\rm min, \Delta}}$. The result for the leading large--$b$
behavior is quoted in Eq.(\ref{f_pi_Delta_asymp}).  Note also that the
representation (\ref{f_pi_Delta_gauss}) can be used for direct
numerical evaluation of $f_{\pi\Delta} (y, b)$.
\end{document}